\documentclass[journal=jacsat,manuscript=article]{achemso}

\usepackage{chemformula} 
\usepackage[T1]{fontenc} 

\usepackage{graphicx}
\usepackage{color}
\definecolor{lightgrey}{rgb}{0.8627,0.8627,0.8627}
\usepackage{amsmath,amssymb}
\usepackage{booktabs}
\usepackage{colortbl}

\newcommand{\ds}{\displaystyle}

\def\<{\langle}
\def\>{\rangle}

\author{Avinash Vicholous Dass}
\affiliation{Centre de Biophysique Mol\'eculaire, CNRS-UPR4301,Rue C. Sadron, 45071, Orl\'eans, France}
\alsoaffiliation{Department of Physics, Ludwig Maximilians University, Amalienstr. 54; D-80799 München, Germany}
\author{Thomas Georgelin}
\affiliation{Centre de Biophysique Mol\'eculaire, CNRS-UPR4301,Rue C. Sadron, 45071, Orl\'eans, France}
\alsoaffiliation{Laboratoire de Réactivité de Surface, UMR 7197, Sorbonne Université, Paris, France.}
\author{Frances Westall}
\affiliation{Centre de Biophysique Mol\'eculaire, CNRS-UPR4301,Rue C. Sadron, 45071, Orl\'eans, France}
\author{Frédéric Foucher}
\affiliation{Centre de Biophysique Mol\'eculaire, CNRS-UPR4301,Rue C. Sadron, 45071, Orl\'eans, France}
\author{Paolo De Los Rios}
\affiliation{Institute of Physics, School of Basic Sciences, Ecole Polytechnique F\'ed\'erale de Lausanne - EPFL,
Lausanne, CH-1015, Switzerland}
\alsoaffiliation{Institute of Bioengineering, School of Life Sciences, 
Ecole Polytechnique F\'ed\'erale de Lausanne, - EPFL, Lausanne, CH-1015, Switzerland}
\author{Daniel Maria Busiello}
\affiliation{Institute of Physics, School of Basic Sciences, Ecole Polytechnique F\'ed\'erale de Lausanne - EPFL,
Lausanne, CH-1015, Switzerland}
\author{Shiling Liang}
\affiliation{Institute of Physics, School of Basic Sciences, Ecole Polytechnique F\'ed\'erale de Lausanne - EPFL,
Lausanne, CH-1015, Switzerland}
\author{Francesco Piazza}
\affiliation{Universit\'e d'Orl\'eans, UFR CoST Sciences et Techniques,
1 rue de Chartres, 45100 Orl\'eans, France}
\alsoaffiliation{Centre de Biophysique Mol\'eculaire, CNRS-UPR4301,Rue C. Sadron, 45071, Orl\'eans, France}
\email{Francesco.Piazza@cnrs-orleans.fr}

\title[\small ]
{Equilibrium and non-equilibrium furanose selection in the 
ribose isomerisation network} 

\abbreviations{}

\begin{document}

\begin{abstract}
\noindent The exclusive presence of $\beta$-D-ribofuranose
in nucleic acids is still a conundrum in prebiotic chemistry,
given that pyranose species are substantially more stable at equilibrium.
However, a precise characterisation of the relative furanose/pyranose fraction
at temperatures higher than about 50$^{\,\rm o}$C is still lacking.
Here, we employ a combination of  NMR measurements and  statistical
mechanics modelling to predict a population inversion
between furanose and pyranose at equilibrium at high temperatures. \\
\indent More importantly, we show that a steady temperature gradient
may steer an open isomerisation network into a non-equilibrium steady state
where furanose is boosted beyond the limits set by equilibrium thermodynamics.
Moreover, we demonstrate that nonequilibrium selection of furanose is maximum at optimal dissipation,
as gauged by the temperature gradient and energy barriers for isomerisation.
The predicted optimum is compatible with temperature drops
found in hydrothermal vents associated with extremely fresh lava flows on the seafloor.
\end{abstract}

\section{Introduction}
                     
Ribose plays a central role in the chemistry of modern life on Earth.
The sugar backbones of ribonucleic acids (RNA) and deoxyribonucleic acids (DNA) 
are formed by polymeric chains of nucleotides which contain a ribose and a deoxyribose 
(a cyclic pentose without a 2$^\prime$ oxygen), respectively. Moreover, ribose
is a key component in modern metabolism, notably through ATP, which also incorporates 
its 5-C anomer, i.e. ribofuranose.
Therefore, stabilisation and reactivity of ribose formation is a critical
topic in prebiotic chemistry and origins of life studies~\cite{Lazcano:1996aa}\,.\\
\indent While the relative stability, abundance,
phosphorylation and glycosylation of ribose anomers, and 
D-ribose (C$_5$H$_{10}$O$_5$) in particular, is clearly a
pre-eminent issue in  the prebiotic  scenario of 
so-called ``RNA world''~\cite{Higgs:2015aa,Gilbert:1986aa}\,, it is increasingly 
being recognised that this topic is, more generally, central to the so-called  
``metabolism-first'' (MF) scenario~\cite{Martin:2008aa}\,.
As opposed to the ``gene-first'' scenario,  MF  hypotheses 
suggest that life originated from energetically driven geochemical networks that
evolved under strong non-equilibrium thermodynamical constraints,
eventually leading to the emergence of conserved metabolic 
pathways at the core of modern biochemical cycles~\cite{Martin:2008aa}\,. 
Of particular relevance in this context  appears the 
role of RNA in scenarios involving collective autocatalytic sets, i.e.
autocatalytic ensembles of molecules that can  
reproduce\cite{Kauffman:1986aa}\,, process and assimilate diversified substrates~\cite{Arsene:2018aa} or, 
more generally, autocatalytic RNA sets 
that demonstrate spontaneous self-assembly~\cite{Vaidya:2012aa}\,.\\
\indent As is the case with all carbohydrates, 
ribose might have been synthesised by polymerisation of 
formaldehyde in abiotic conditions from the
formose reaction~\cite{Butlerow:1861ab,Breslow:1959aa,Orgel:2000aa}\,, 
although this reaction has so far only yielded 
ribose in very minor quantities~\cite{Kopetzki:2011aa}\,. 
After synthesis the ribose molecule is unstable in aqueous solution~\cite{Larralde:1995aa}\,. 
For example, at pH 7 and 90\,$^{\rm \,o}$C, its half-life 
is about 10 hours~\cite{Georgelin:2015aa}\,. Despite its lack of stability, ribose is the 
exclusive constituent of the carbohydrate 
backbone of RNA in its $\beta$-D-ribofuranose form (hereafter simply furanose).  
At thermal equilibrium, at room temperature and ambient pressure, this enantiomer represents only 12 \% 
of all ribose molecules in solution~\cite{Drew:1998aa}\,. 
Thus, it is of paramount importance to explain the exclusive 
incorporation 
of $\beta$-D-ribofuranose in RNA. There are many potential causes for this,  ranging from incorporation 
due to a specific chemical process 
during phosphorylation or glycosylation, or perhaps due to specific physicochemical conditions on the early 
Earth that led to a significant increase in the proportion of furanose. \\
\indent The phosphorylation of ribose could favour a specific
enantiomer. It has been shown that the glycosylation of ribose under the $\alpha$-furanose 
form can lead to the formation of $\beta$-furanose nucleosides~\cite{Singh:2014aa}\,. 
In order to solve the problems of stability and  enantiomeric structure, some studies 
have analysed the ability of ribose to 
form borate or silicate complexes that are more stable in solution~\cite{Kolb:2004aa,Benner:2010aa}\,. 
Furthermore, theoretical studies have also 
shown that silicate/ribose complexes would be formed exclusively from the furanose form because, with this structure, 
the HO-C-C-OH dihedral angle is sufficiently small to allow the formation of a planar 
five-membered ring~\cite{Lambert:2004aa}\,. The silicate or borate 
scenarios have shown the high potential of inorganic/organic interactions, although the presence of 
significant borate on the early 
Earth is unclear~\cite{Westall:2018aa}\,.  
Moreover, since coordination processes seem to have an impact on isomerisation, it is 
possible that thermal effects 
at equilibrium could also have an impact on these processes and on the anomeric ratios. Thus, it is
important to investigate 
the effect of temperature on ribose isomerisation at thermal equilibrium. This  aspect has not 
yet been thoroughly 
studied, despite the fact that temperatures on the primitive Earth, at least at the rock/water 
interface where prebiotic 
reactions were taking place, 
were likely higher than 50$^{\rm \,o}$C~\cite{Westall:2018aa,Tartese:2017aa}\,. 
It is to be noted, however, that the environments where prebiotic chemical reactions 
thrived~\cite{Westall:2018aa,Tartese:2017aa,Smith:2016aa}\,, 
such as hydrothermal 
vents and their immediate vicinity, were by no means at thermal equilibrium.  
Most probably, chemical reaction networks proceded under the action of high, 
steady gradients of temperature,
pH~\cite{Moller:2017aa} and chemical activity of key molecular species, such as 
water~\cite{Smith:2016aa,Westall:2018aa,Mast:2013aa,Agerschou:2017ab,Lane:2017aa,Morasch:2016aa}\,. \\
\indent It is well known that open chemical systems~\cite{Rao:2016aa} driven far from equilibrium  
may settle in non-equilibrium steady states (NESS) that  bear little resemblance 
with equilibrium ones~\cite{Nicolis:1977aa}\,.
Well-known examples are regulatory and metabolic network in biochemistry~\cite{Fang:2019aa} 
and, more generally, all chemical transformations that proceed thanks to catalysts, such 
as enzyme catalysis in biology,  where substrate and product species are kept at fixed concentration 
(chemostatted) by an external source of energy~\cite{Stich:2016aa}\,. 
The work done on the system to enforce the required 
chemical potential difference leads to sustained dissipative currents that steer the chemical transformations
away from equilibrium~\cite{Ge:2012aa}\,. In more chemical terms, a NESS should be considered as a regime of
sustained kinetic control, with reference to the well-known concept of transient kinetic control 
of chemical reactions, as opposed to so-called thermodynamic control regime, that is,
thermodynamic equilibrium~\cite{Pross:2003aa}\,. More recently, the subtle effects of steady temperature 
gradients on chemical reaction networks have been brought to the fore, especially as regards 
the coupling of sustained mass currents in physical space and chemical currents 
in state space, which, based on  kinetic rules, may push the steady-state molar fraction of certain 
chemical species away from equilibrium values~\cite{Busiello:2019aa,Roduner:2016aa}\,.\\
\indent In this paper, we focus on a coarse-grained chemical reaction network describing 
D-ribose isomerisation. Our main working hypothesis is that $\beta$-furanose could be 
ultra-stabilised under steady non-equilibrium 
conditions beyond the limits imposed by equilibrium thermodynamics.
In order to investigate this idea and quantify the necessary conditions in terms of the 
unknown kinetic parameters, the first step was to accurately characterise 
the anomeric ratios at equilibrium as a function of temperature through NMR.  
The data are fitted to a simple equilibrium  model, solved under the constraint of detailed balance. 
The first  finding is that, in view of the large entropic degeneracy 
associated with sub-conformations of  $\alpha-$ and $\beta-$furanose, 
the populations of these high-energy species are predicted 
to increase with  temperature, eventually leading to a population 
inversion at high temperature. \\
\indent Equipped with the thermodynamic parameters derived from our experiments, 
in the second part of the paper we report a theoretical exploration of D-ribose isomerisation 
under the action of an 
applied steady gradient of temperature. In this setting, we show that mass currents sustained 
by the temperature gradient can 
couple to chemical transformation steps and drive the system into a steady state where the most 
unstable furanose species can be stabilised beyond the equilibrium limits. 
The crucial parameters that regulate this effect are (i) energy barriers for chemical transformations and 
(ii) the Damk\"ohler number, i.e. the non-dimensional ratio between characteristic 
chemical and mass transport rates.

\section{NMR characterisation of D-ribose isomerisation at thermal equilibrium} 

In order to investigate the  behaviour of ribose isomerisation in solution at equilibrium
at increasing temperatures, 
we carried out $^{13}$C NMR experiments. This quantitative technique allows  
unambiguous identification of each anomeric form of ribose by studying the C1 signal of ribose
(see supplementary information for more details).
In the following, we denote with $\alpha$F,$\beta$F and $\alpha$P,$\beta$P
the two enantiomers of furanose (F) and pyranose (P), respectively.
We recorded two different sets of spectra.
The first set was collected in the temperature range 10-80$^{\rm o}$C  
in pure water. A second set of measurements were conducted in the temperature range 10-25$^{\rm o}$C 
in simulated Hadean sea water~\cite{Dass:2018aa}\,, with the purpose  
of investigating isomerisation at equilibrium in the presence of relevant saline conditions.
The composition of Hadean oceans was slightly different to that of modern seawater, 
particularly having lower SO$_4$, being  saturated in silica, having higher Fe, Mg, and Ca 
as well as other elements and molecules related to abundant hydrothermal activity in 
an ultramafic crust (see supplementary material for more information).\\
%
\begin{figure}[!t]
\centering
\includegraphics [width=7truecm]{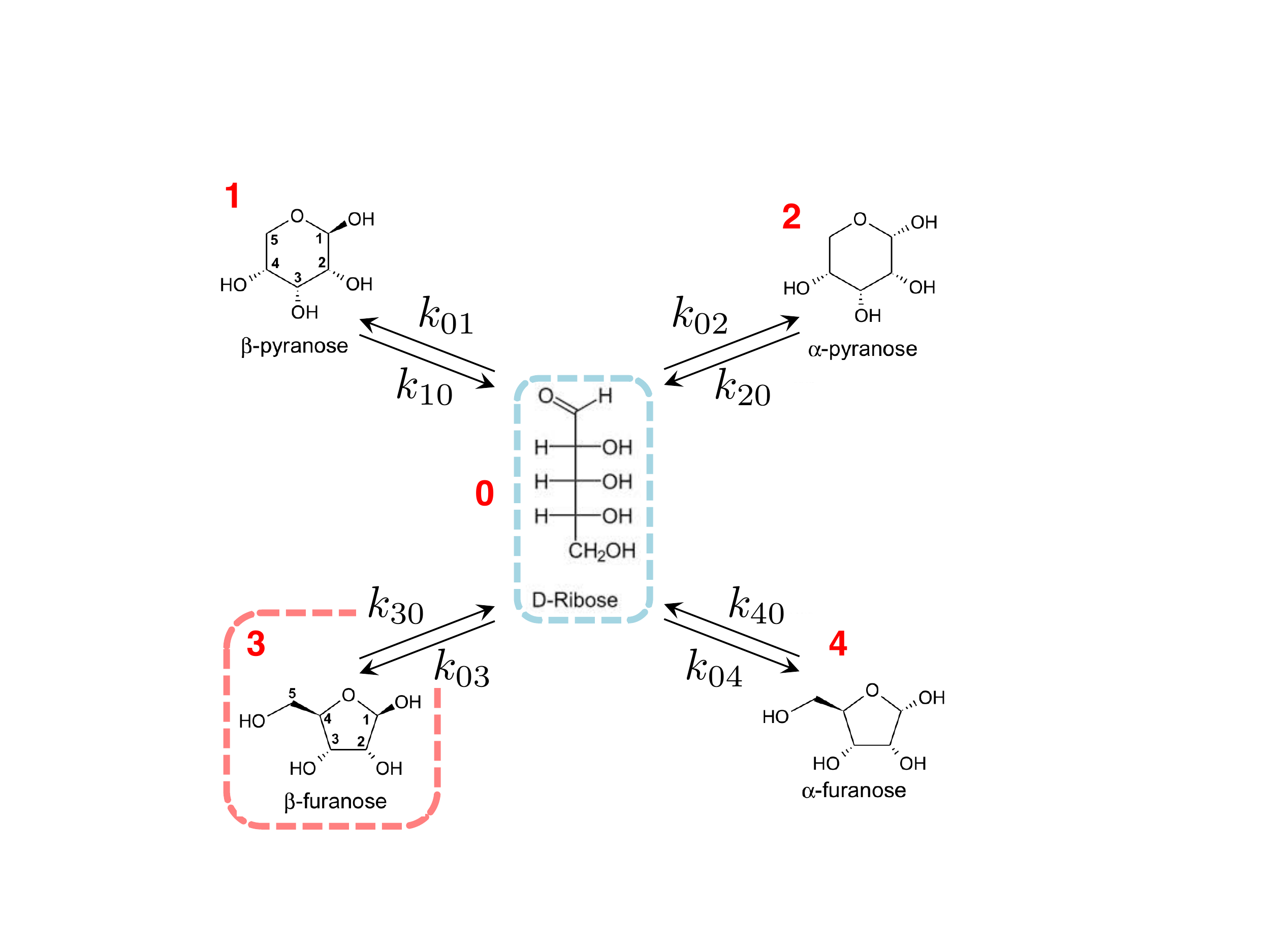}
\hspace{1mm}
\includegraphics [width=7truecm]{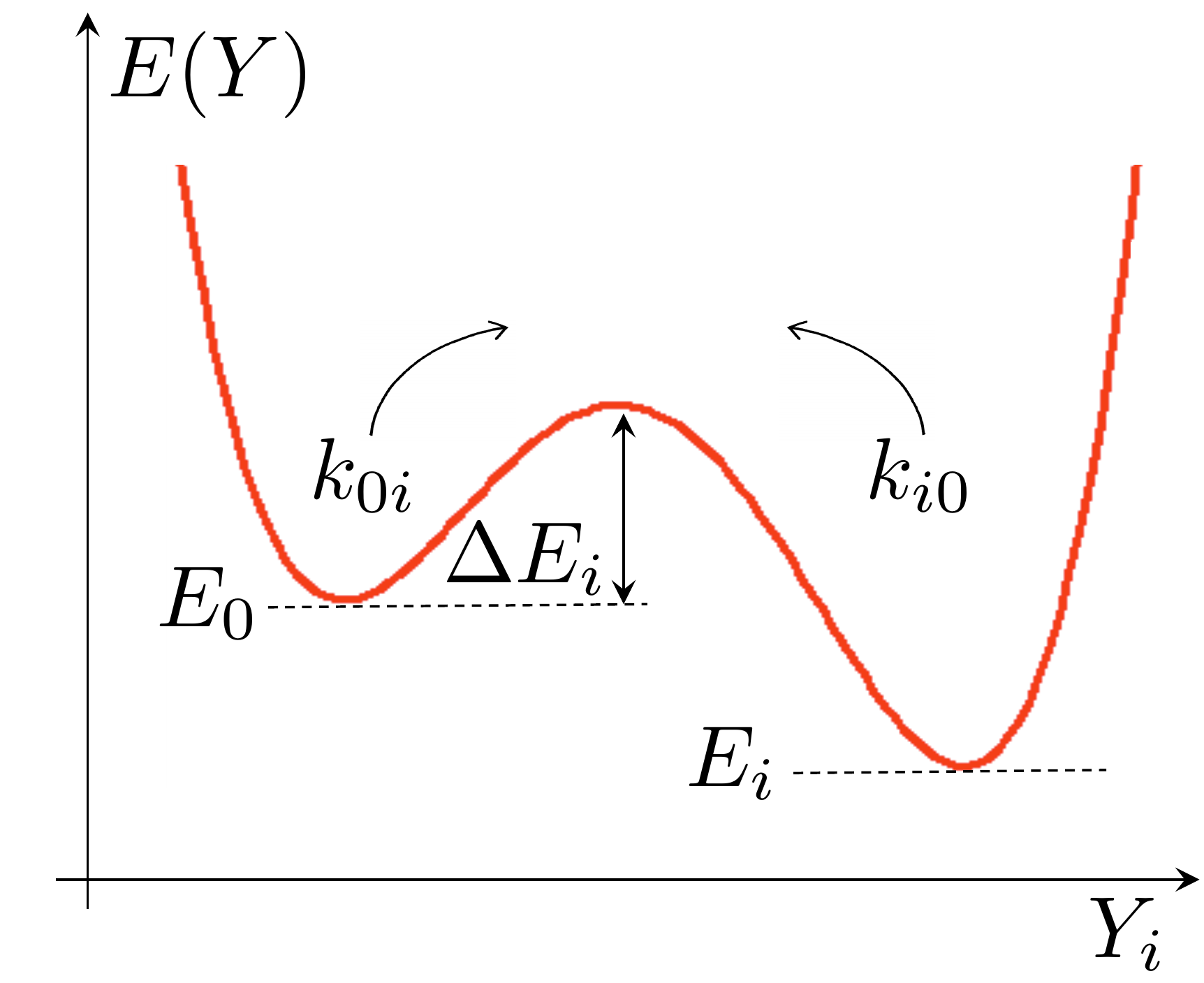}
\caption{{\bf Scheme of the D-ribose isomerisation reaction network and model energy landscapes}. 
The two pyranoses ($\alpha$ and $\beta$) and 
the two furanoses ($\alpha$ and $\beta$) are in equilibrium at a given temperature and pressure
with the high-energy linear conformation. Right: scheme of the Gibbs free energy landscape along 
a representative 
reaction coordinate for the transition between the linear chain and the $i$-th anomer. The 
energy of the linear conformation is the highest, i.e. $E_0 > E_i \ \forall \, i$~\cite{Cocinero:2012aa}\,, 
corresponding to a population that is undetectable in solution via NMR measurements.}
\label{f:reactn}
\end{figure}
%
%
\indent The thermal behaviour of D-ribose isomerisation at the resolution of our NMR 
experiments is governed by the energy (enthalpy) differences 
between different states, as well as by the degeneracies associated with 
the different sub-structures of the four main conformations~\cite{Cocinero:2012aa}\,, which 
cannot be resolved in our spectra. Based on these considerations, we can formulate 
a simple equilibrium model as depicted in Fig.~\ref{f:reactn} (left panel), that
involves the linear conformation and the four ring species, $\beta$P, $\alpha$P, $\beta$F, $\alpha$F. 
These are labelled with integers from $0$ to $4$, respectively, as illustrated 
in Fig.~\ref{f:reactn}.
According to the above considerations, the transition rates can be written 
as the product between a temperature-independent 
geometric (i.e. entropic) velocity and an Arrhenius-like term, gauging the thermal activation of the 
transition.  Let us denote with $x_i$ ($i=0,1,\dots,4$) the relative concentration (molar fraction) of 
the $i$-th species, so that $\sum_{m=0}^4 x_m=1$. 
With reference to Fig.~\ref{f:reactn} (right panel), we may thus write
\begin{equation}
\label{e:rates}
\begin{cases}
k_{0i} = k_{0i}^\infty e^{-\beta \Delta E_i} & \\
k_{i0} = k_{i0}^\infty e^{-\beta (\Delta E_i + E_0 - E_i)}                                          
\end{cases}
\end{equation}
where $\beta^{-1} = k_BT$, $k_B$ being Boltzmann's constant. 
The energy of the open linear conformation is the highest, averaging about 19 kJ/mol in the gas  
phase~\cite{Cocinero:2012aa}\,. This is confirmed by our solution measurements, where 
this conformation is undetectable (see supplementary information). 
In the limit  $\beta E_0 \gg 1$
the stationary solution of the rate equations for the ribose 
isomerisation network depicted in Fig.~\ref{f:reactn} at thermal equilibrium read
\begin{equation}
\label{e:xeqasy}
x_i = \frac{\eta_i e^{-\beta E_i}}
           {\sum_{m=1}^4 \eta_m e^{-\beta E_m}}  \qquad   i=1,2,\dots,4                       
\end{equation}
where $\eta_i = k_{0i}^\infty/k_{i0}^\infty$.
The degeneracy factors $\eta_i$ embody the entropic contributions to the interconversion rates,
so that the difference in entropy between states $i$ and $j$ can be computed as 
$\Delta S_{ij} = k_B \log(\eta_i/\eta_j)$.
These are mainly associated with the specificities of the multidimensional free-energy landscapes,
including local minima corresponding to sub-conformations of each anomer that are 
in fast equilibrium with the main isoforms corresponding to the four detected NMR lines.\\
%
\begin{figure}[ht!]
\centering
\includegraphics [width=12truecm]{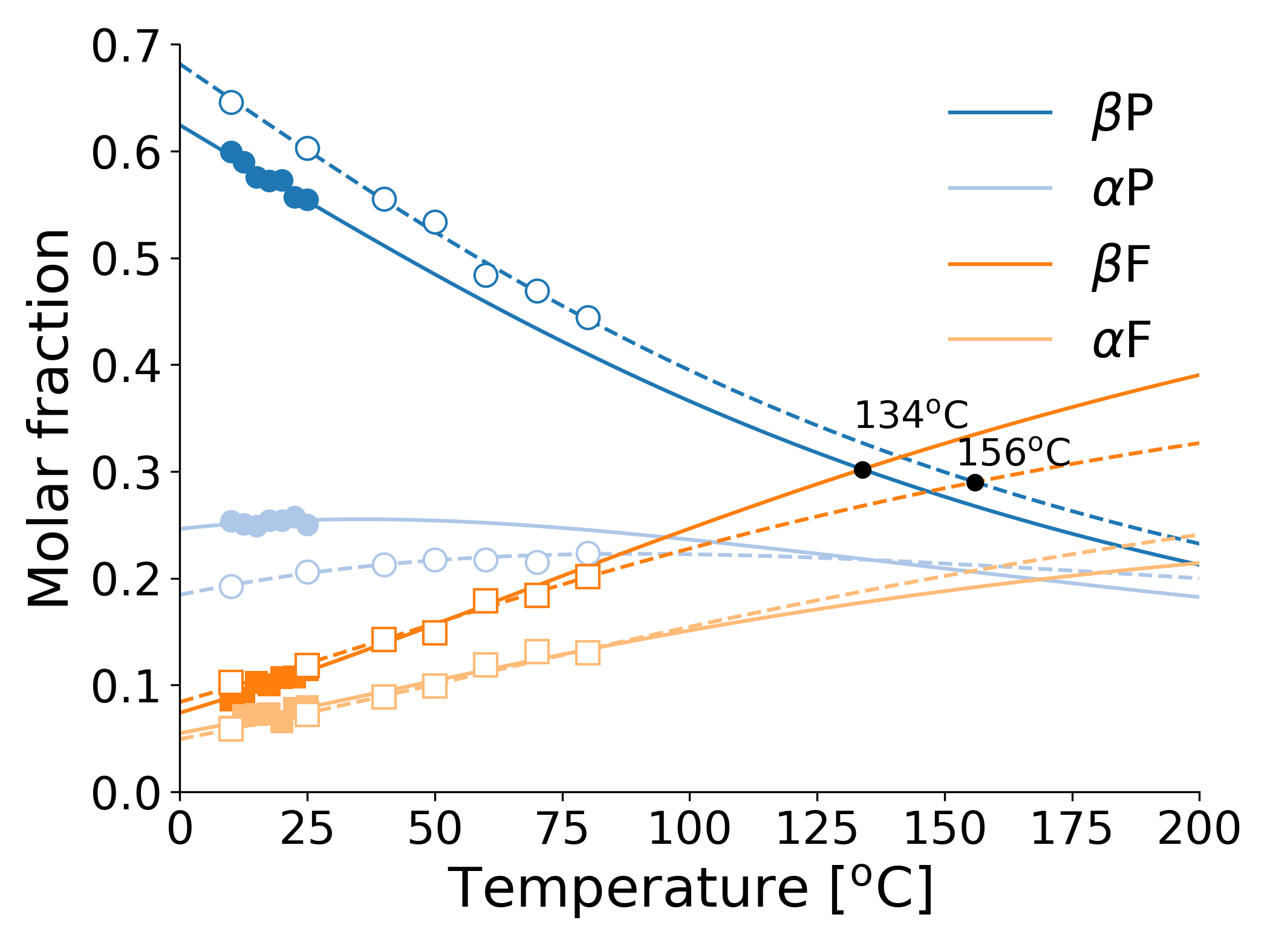}
\caption{{\bf Population inversions in D-ribose isomerisation network 
at thermal equilibrium.}
The average molar fractions of the four main anomers of D-ribose are plotted
versus temperature for the experiments performed in pure water (open symbols) and
in Hadean model water (filled symbols, see supplementary material for more details). 
Lines are fits performed simultaneously with Eqs.~\eqref{e:xeqasy} 
to the four anomer temperature  series  (see Table~\ref{t:eqpar}). 
Dashed lines: pure water, solid lines: Hadean model water.
In practice, we minimised a single cost function that included a total 
of $4N_T$ points, where $N_T$ is the number of temperature points considered 
($N_T=7$ for both pure and Hadean water).}
\label{f:NMReq}
\end{figure}
%
%
%
\begin{table}[t!]
\begin{center}
\begin{tabular}{@{\extracolsep{\fill}}l r r r r r r }
\cmidrule{1-7}
&\multicolumn{3}{c}{Pure water} & \multicolumn{3}{c}{Hadean water} \\
\cmidrule{2-4}\cmidrule{5-7}
Species & $E_i$ [kJ/mol] & $\eta_i$ &  $x^\infty_i$ &   $E_i$ [kJ/mol] & $\eta_i$ & $x^\infty_i$ \\
\cmidrule{1-7}
1, $\beta-$pyranose  &   \verb"0"     &  \verb"0.75"  & \verb"0.01" & \verb"0"    & \verb"0.39"  & \verb"0.01"   \\
2, $\alpha-$pyranose &   \verb"6.2"   &  \verb"3.14"  & \verb"0.05" &  \verb"4.2"  & \verb"0.96"  & \verb"0.02"  \\
3, $\beta-$furanose  &   \verb"13.0"  &  \verb"29.27" & \verb"0.47" & \verb"14.7" & \verb"29.88" & \verb"0.71"   \\
4, $\alpha-$furanose &   \verb"14.2"  &  \verb"29.41" & \verb"0.47" &  \verb"13.1" & \verb"10.88" & \verb"0.26"  \\
\cmidrule{1-7}
\end{tabular}
\end{center}
\caption{\label{t:eqpar} Thermodynamic parameters describing the equilibrium 
of D-ribose isomerisation in solution at fixed temperature and pressure as estimated 
by fitting the equilibrium theory~\eqref{e:xeqasy} to $^{13}$C  NMR data.}
\end{table}
%
\indent Fig.~\ref{f:NMReq} shows clearly that the simple equilibrium model~\eqref{e:xeqasy}
provides an excellent description of the NMR data as temperature increases.   
The first observation to be made is that the two high-energy furanose species become 
stabilised as temperature increases, while the equilibrium 
molar fractions of the more stable pyranoses decrease correspondingly. As a consequence, 
our analysis predicts that a series of population inversions between the high-energy species 
$\beta-$furanose and  the two low-energy ones, $\alpha-$pyranose, 
should occur upon increasing the temperature.
The first of these should occur at a temperature of about 
$k_B^{-1}(E_3-E_2)/ \log(\eta_3/\eta_2) \approx 93.2$ $^{\rm o}$C, 
while the $\beta-$furanose-$\beta-$pyranose inversion is predicted to occur at a temperature
$k_B^{-1}(E_3-E_1)/\log(\eta_3/\eta_1) \approx 156$ $^{\rm o}$C.
Interestingly,  our calculations place the first furanose-pyranose inversion at a  
temperature some $20$ $^{\rm o}$C  lower 
in Hadean water, at about $134$  $^{\rm o}$C (see Fig.~\ref{f:NMReq}).\\
\indent From a thermodynamic point of view, the population inversions are explicitly 
related to the large entropic degeneracy associated with the furanose states. 
The infinite-temperature molar fractions $x_i^\infty$ provide a clear 
illustration of this trend~\footnote{Even if, of course, the molecular species examined 
are certainly not stable beyond a certain temperature.}. From Eq.~\eqref{e:xeqasy}, these
read 
\begin{equation}
\label{e:xinf}
x^\infty_i = \frac{\eta_i}
           {\sum_{m=1}^4 \eta_m}  \qquad   i=1,2,\dots,4                       
\end{equation}
As illustrated by the calculations reported  in Table~\ref{t:eqpar}, equilibrium thermodynamics
predicts that $\beta$-furanose will dominate at high temperatures. Its stability under 
given geochemical conditions is thus the only limitation to the maximum fraction of 
furanose that can be produced at thermal equilibrium by increasing the temperature. Interestingly, this 
effect is magnified in Hadean water, an environment where 
temperature-induced  boosting of $\beta$-furanose at equilibrium appears to have been  easier
(see Table~\ref{t:eqpar}). 
%

\section{The furanose population can be boosted beyond equilibrium in a steady temperature gradient}

\noindent Our NMR experiments have made it very clear that $\beta$-furanose is progressively 
stabilised at increasing temperature at thermal equilibrium. However, 
it is well known that typical prebiotic environments, such as hydrothermal vents and the adjacent porous 
sediments and chemical precipitates,
were characterised by strongly non-equilibrium conditions, such as 
steady gradients of temperature, pH and water 
activity~\cite{Barge:2019aa,Westall:2018aa,Sojo:2016aa,Spitzer:2013aa,Cockell:2006aa}\,.
Intriguingly, in complex systems with multiple states,  
it has been shown that the rate of dissipation 
(equivalently, the rate of entropy production)  
conveys key information on the {\em selection} of states  that are
favoured away from equilibrium~\cite{Endres:2017aa,England:2015aa,Meysman:2010aa}\,. 
In more chemical terms, a given reaction network driven far from equilibrium 
is placed under a state of sustained kinetic control.
In such conditions, the energy barriers that set the velocity of chemical transformations, which are
irrelevant at equilibrium,
become key in selecting the steady-state populations.\\
\indent Taken together, in the context of D-ribose isomerisation kinetics, 
the above considerations prompt the question whether high-energy furanose species 
could be further stabilised under steady non-equilibrium conditions, such as an
imposed temperature gradient, 
beyond the limits imposed by  equilibrium thermodynamics. 
The in-depth, quantitative analysis of the thermodynamic 
equilibrium properties performed in the first part of this study enables
us to investigate this question from the vantage point of a  
model parameterised on solid experimental evidence.\\
%
\begin{figure}[t!]
\includegraphics [width=16truecm]{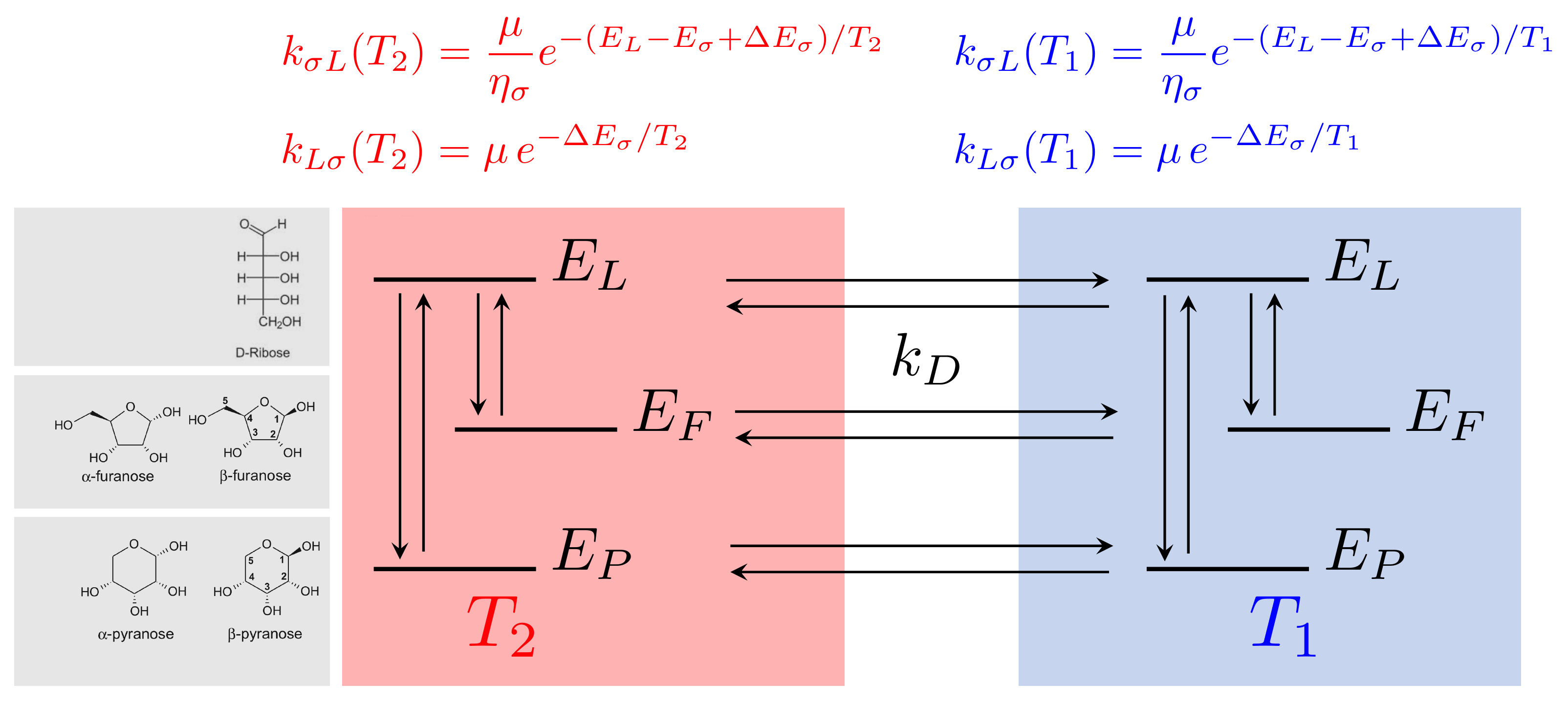} 
\caption{{\bf Simple model of  ribose isomerisation  in a steady temperature 
gradient ($T_2 > T_1$)}. The highest-energy linear species has energy $E_L$, $\alpha$F and $\beta$F 
have been coalesced into the high-energy state $F$ (energy $E_F$),
while $\alpha$P and $\beta$P are coarse-grained into the ground state (and reference energy), 
$E_P$. Chemical rates are expressed as the product of a  velocity, $\mu$, and an Arrhenius 
term that depends on the energy barriers $\Delta E_\sigma$ ($\sigma=F,P$) and include
entropic degeneracy factors, $\eta_\sigma$, in accordance with our experimental findings.}
\label{f:3statesmodel} 
\end{figure}
%
%
\indent As proposed in a recent work~\cite{Busiello:2019aa}\,, a simple 
way to examine a chemical reaction network under the 
action of a steady temperature gradient is to replicate the
same system in two separate compartments, each thermalised at a different temperature, 
and able to exchange reactants as dictated by specific transport rates (see Fig.~\ref{f:3statesmodel}).
Furthermore, with no loss of generality,  we can coarse-grain the kinetics 
by letting the two pyranoses coalesce into a single low-energy species, $E_P$, and likewise reunite the 
two furanoses in a single high-energy moiety, $E_F$. 
The third species represents the open chain --  the high-energy transition state, $E_L$.
For simplicity,  mass exchange  (e.g. convection, diffusion) is described by a 
single transport rate, $k_D$. \\
\indent In a system where chemical reactions and transport are coupled, 
the central physical parameter is the ratio of the respective characteristic rates. 
If $D$ denotes the typical molecular diffusion coefficient of reactants, 
$\ell$ some relevant dimension of interest and $\mu$ the typical velocity of 
chemical transformations (see Fig.~\ref{f:3statesmodel}), then the key parameter is known as the
Damk\"{o}hler number~\cite{Goppel:2016aa}\,, Da = $\mu \ell^2/D = \mu/k_D$. 
This parameter measures the relative time scale of kinetic effects with respect to 
mass transport and thus offers a single expedient  gauge for the degree 
of coupling between the kinetics of chemical transformations
and the transport of reactants and products. In the fast reaction limit,  Da $\gg 1$,
transport is not swift enough to couple molecular species that undergo chemical transformations 
in separate compartments. Therefore, in the stationary state each box settles at thermal 
equilibrium at its own temperature. \\
\indent  When transport becomes faster, as happens for example in the 
presence of strong convective currents in the vicinity of hydrothermal vents~\cite{Haase:2007aa}, 
exchange of mass between the two compartments will eventually approach
timescales comparable to chemical transformations. 
In this regime, the steady populations of reactants are no longer 
dictated by detailed balance. Rather, they are governed by a kinetic selection mechanism that
depends on the relative values of the energy barriers, $\Delta E_F, \Delta E_P$ (see 
Fig.~\ref{f:reactn} and Fig.~\ref{f:3statesmodel}) and the strength of the temperature gradient, $\Delta T$. 
Simply put, 
the molecular species that are generated most quickly from the transition state, $E_L$,
will be boosted beyond the equilibrium population
by large dissipative mass currents sustained by the temperature gradient. 
Remarkably, provided the difference between the barriers is large enough, this
is true irrespective of whether  the fastest state is the least or the most stable at equilibrium.
Figs.~\ref{f:PPeq}a and~\ref{f:PPeq}b illustrate the scenario where the imposed temperature gradient 
leads to non-equilibrium boosting of the high-energy furanose species. \\
\indent Interestingly, while the extent of non equilibrium stabilisation of furanose 
in terms of its excess population with respect to equilibrium for $k_D \gg \mu$ (Da $\ll 1)$
is controlled essentially by the temperature gradient, the crossover to the 
fast-transport limit, Da $<$ Da$^\ast$, is also governed by kinetic parameters. This
can be encapsulated in a remarkably transparent formula (see supplementary material), namely
\begin{equation}
\label{e:k2sTHE}
\text{Da}^\ast = \frac{1}{\ds e^{-\Delta E_F/k_BT_M} + e^{-\Delta E_P/k_BT_M} }                
\end{equation}
where $T_M = (T_1 + T_2)/2 = T_1 + \Delta T/2$ is the average temperature  of the system.
From Eq.~\eqref{e:k2sTHE} it can be readily seen that the requirement for kinetic selection, i.e. 
scenarios where one of the barriers is appreciably higher than the average temperature
and the other lower,  correspond to the timescale-matching condition Da$^\ast = \mathcal{O}(1)$
when the fast reaction is regulated by an energy barrier of roughly 
the same order as the average temperature. \\
%
\begin{figure}[t!]
\begin{tabular}{ccc}
\includegraphics[width=7.truecm]{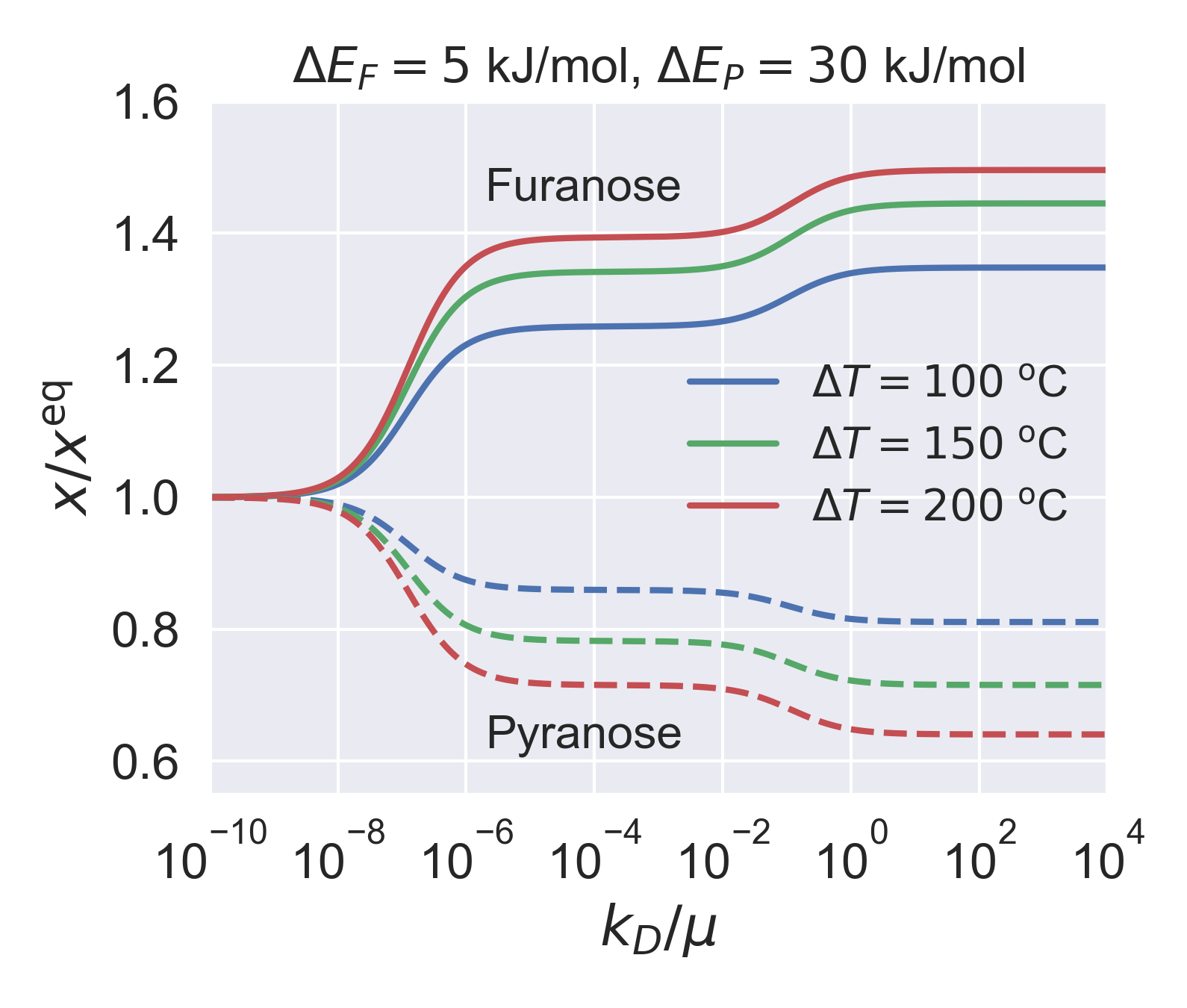} &&
\includegraphics[width=7.truecm]{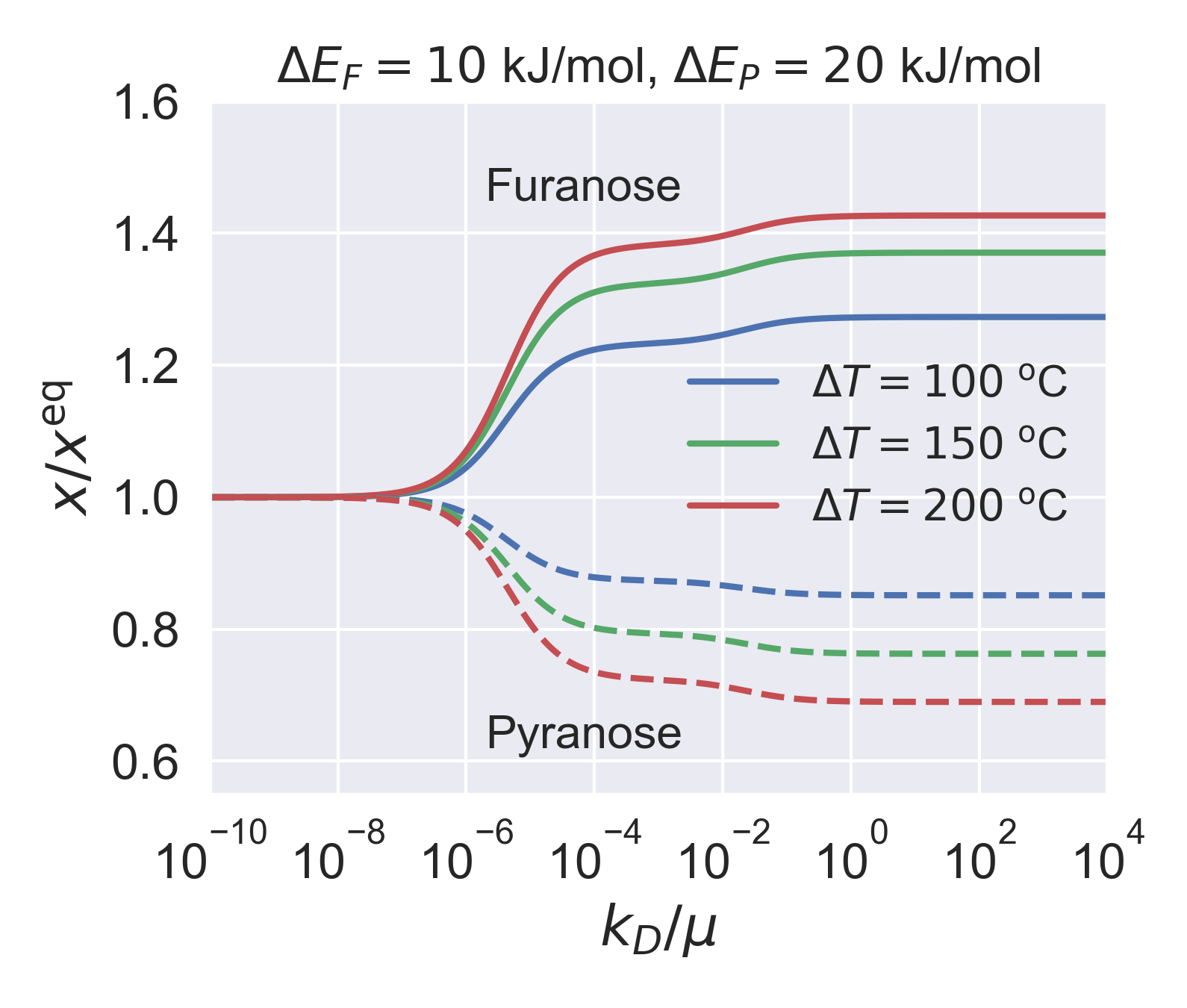} \\
(a) && (b)
\end{tabular}
\begin{tabular}{c}
\includegraphics [width=10truecm]{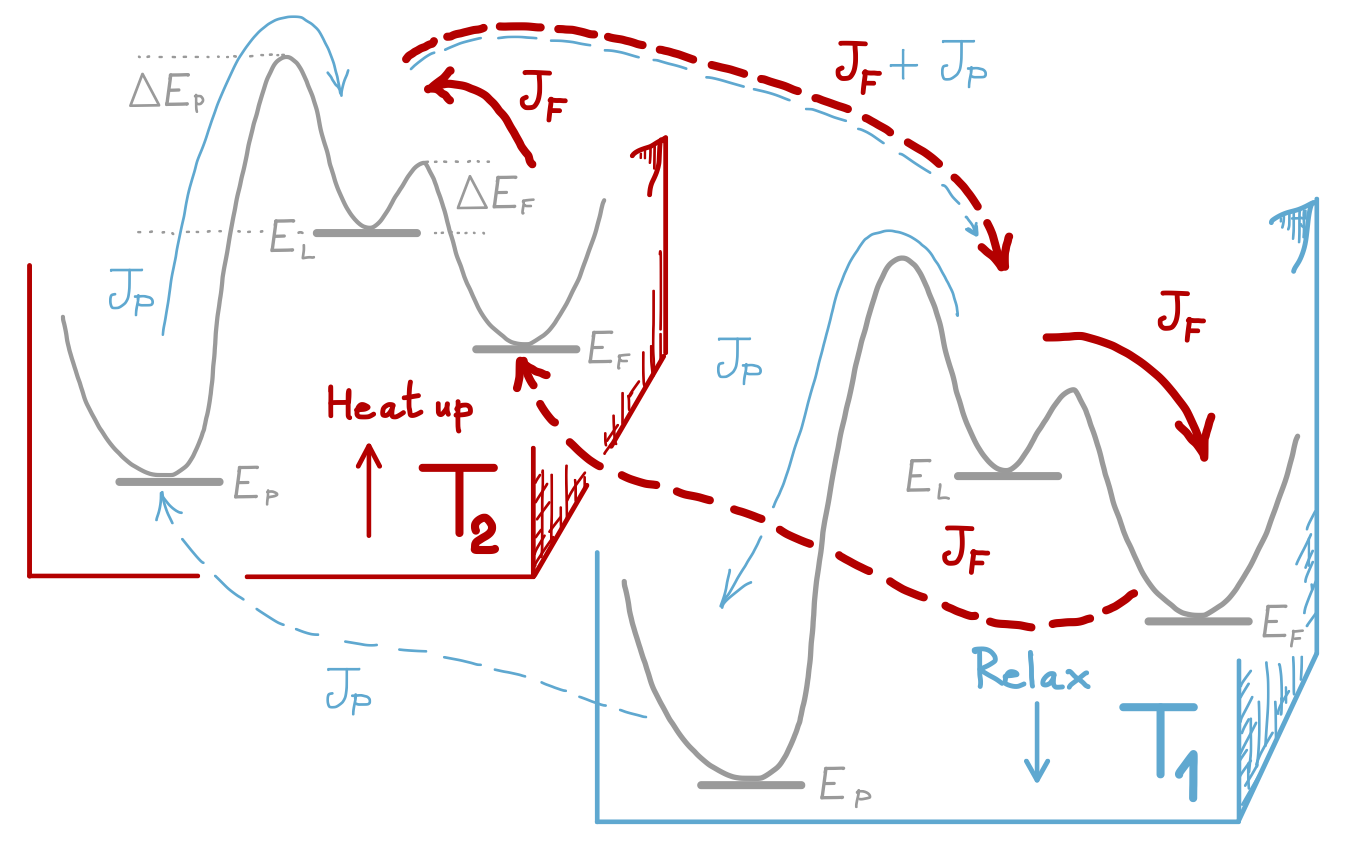}\\
 (c) 
\end{tabular}
\caption{{\bf Top. Increasing the transport rate for $\Delta E_F < \Delta E_P$ leads
to nonequilibrium kinetic selection of furanose beyond thermodynamic equilibrium.}
The steady-state molar fractions of furanose (solid lines) and pyranose (dashed lines)
computed for the two-box model illustrated in Fig.~\ref{f:3statesmodel} 
are plotted against the transport rate for different choices of the temperature 
gradient $\Delta T = T_2 - T_1$ and two choices of energy barriers
(see Supplementary Material for the mathematical details). 
Populations are normalised to the equilibrium values, 
i.e. $x^{\rm eq} \equiv (x_F^{\rm eq}(T_1)+x_F^{\rm eq}(T_2))/2$.
{\bf Bottom. The energy harvested at the hot end is dissipated at the cold end, 
thus causing population inversion}.
Illustration of the steady system of currents that sustain the non-equilibrium 
stabilisation of furanose for $\Delta E_P > \Delta E_F$. Dashed arrows
denote mass transport, solid lines stand for chemical transformations.
The current $J_P$ is much smaller than $J_F$ in the fast transport limit, Da $\lessapprox 1$.
Parameters used in the two-box model are: $T_1$ = 60 $^{\rm o}$C,
$\eta_F = 29.34$, $\eta_P = 1.94$, $E_L=19$ kJ/mol, $E_F=13.6$ kJ/mol, 
$E_P=3.1$ kJ/mol, corresponding to the average values for the two furanose and pyranose
enantiomers measured from our equilibrium NMR experiments (see Table~\ref{t:eqpar}).
}
\label{f:PPeq} 
\end{figure}
%
%
\indent The rationale behind non-equilibrium stabilisation of furanose becomes obvious 
once the system of steady fluxes is examined in detail, as illustrated in 
Fig.~\ref{f:PPeq}c (see also Supplementary material).  
When $E_L \to E_P$  relaxation is the faster pathway out of the transition state,   
the furanose population is boosted by a large sustained mass current that 
supplies high-energy linear molecules from the hot end that undergo ring-closure 
at the cold side (thick red arrows). A much smaller 
current circulates in the same direction (thin blue arrows), weakly contributing 
to the steady mass current of linear D-ribose with pyranose molecules that open up at the hot end.\\
\indent To summarise, when time scales for transport and chemical transformations match, 
part of the energy supplied by the temperature gradient 
can be  converted into chemical energy through cycles involving mass transport. 
Thus, non-equilibrium population inversion may occur if the fastest reaction is fast enough, 
as gauged by the relative magnitude  of $\Delta E_P$ and $\Delta E_F$ 
(see Fig.~\ref{f:PPeq}). 
The extent of this non-equilibrium effect is illustrated 
in Table~\ref{t:US}, where the temperature needed to obtain a given
molar fraction of furanose at equilibrium is compared to the gradient required 
to obtain the same value in a non-equilibrium steady state. 
The population boost obtained in a non-equilibrium setting clearly outperforms 
by far the stabilisation 
achievable at equilibrium, allowing for molar fractions of furanose
that could only be produced at exceedingly high temperatures in thermal equilibrium. 
Note also that chemical stability might be a less critical issue far from equilibrium,
as molecules would fly past the hot end transported by large convective currents. 

%
\begin{table}[t!]
\begin{center}
\begin{tabular}{@{\extracolsep{\fill}}l c c c c c c c c }
\toprule
& & & &\multicolumn{2}{l}{$\Delta E_F=5$ kJ/mol}
& \multicolumn{2}{l}{$\Delta E_F=5$ kJ/mol} \\
& & & &\multicolumn{2}{l}{$\Delta E_P=30$ kJ/mol} 
& \multicolumn{2}{l}{$\Delta E_P=20$ kJ/mol} \\
 \cmidrule{5-6}\cmidrule{7-8}
%
%
$T_1$       & $T_2$        & $\Delta T$ & \quad & $x_F$ & $T^\ast$ & $x_F$ & $T^\ast$   \\
\hline\hline
\verb"60"   &  \verb"160"  & \verb"100" & \quad  & \verb"0.47"  & \verb"174"  & \verb"0.46"  & \verb"168" \\
\verb"60"   &  \verb"210"  & \verb"150" & \quad  & \verb"0.56"  & \verb"241"  & \verb"0.55"  & \verb"234" \\
\verb"60"   &  \verb"260"  & \verb"200" & \quad  & \verb"0.62"  & \verb"301"  & \verb"0.61"  & \verb"290" \\
\bottomrule
\end{tabular}
\end{center}
\caption{\label{t:US} Temperature $T^\ast$ required to obtain a given overall molar fraction of 
furanose, $x_F$, at equilibrium, compared to different choices of gradients that yield
the same populations in a non-equilibrium steady state. Two choices of energy barriers are considered.
Note that non-equilibrium stabilisation of furanose is more effective in all cases considered, 
according to the stringent criterion $T^\ast > T_2$.
All temperatures are expressed in $^{\rm o}$C.}
\end{table}

\section{Nonequilibrium selection of furanose is maximum at optimal dissipation} 

\begin{figure}[t!]
\centering
\includegraphics [width=16truecm]{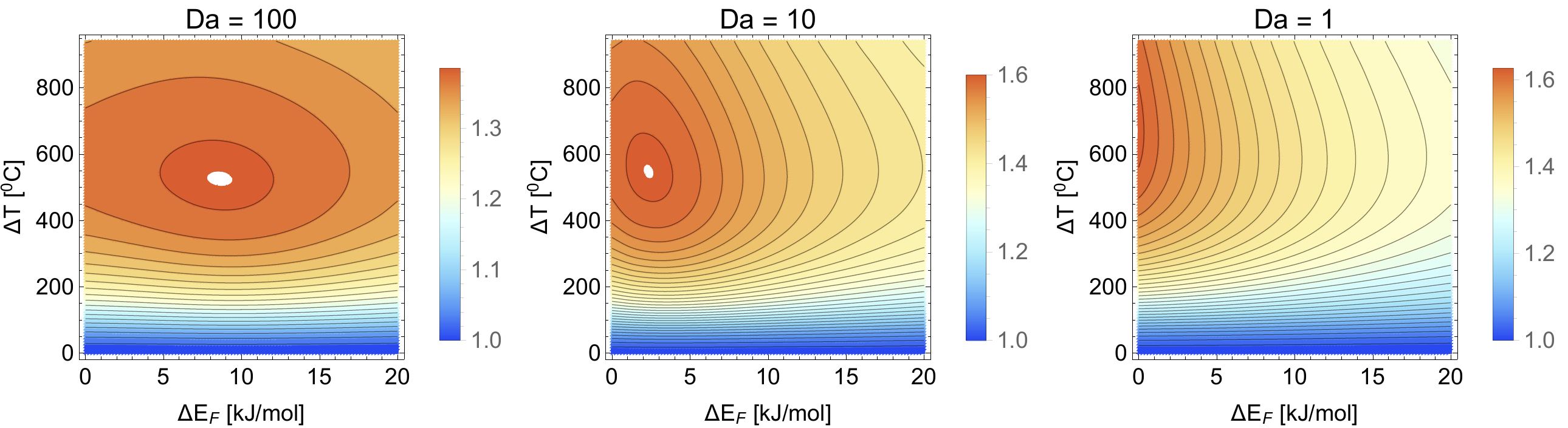}
\includegraphics [width=16truecm]{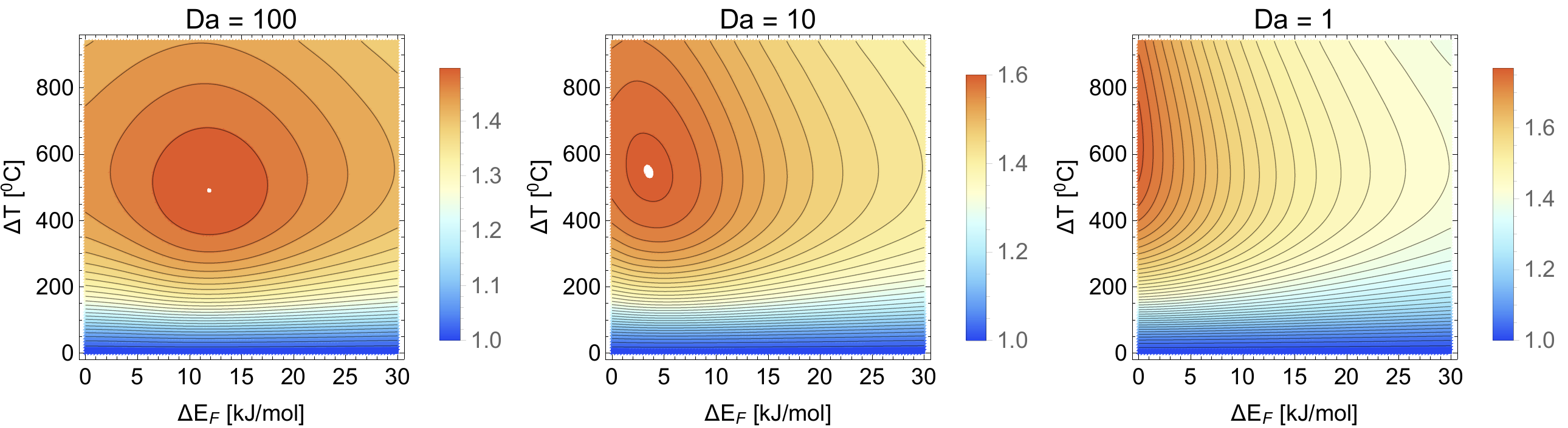}
\caption{{\bf Furanose selection is maximum for specific values of the temperature gradient 
depending on kinetic parameters.} Density plots of the normalised selection indicator $\mathcal{R}_{\rm sel}$, 
Eq.~\eqref{e:Rselcont},
in the $(\Delta T, \Delta E_F)$ plane computed from the steady-state solutions of 
the continuum model~\eqref{e:reactdiff}
for different values of the Damk\"ohler number Da = $\mu/DL^2$. The temperature 
profile is $T(x) = T_1 + \Delta T x/L$. Top panels:
$\Delta E_P = 20$ kJ/mol. Bottom panels: $\Delta E_P = 30$ kJ/mol.
Other parameters as in Fig.~\ref{f:PPeq}.}
\label{f:Rsel} 
\end{figure}

\noindent The simple two-box model discussed in the preceding section can be easily generalised 
in the continuum limit to a system of reaction-diffusion partial differential equations. 
For the sake of simplicity, and with no loss of generality, we shall restrict ourselves 
to a one-dimensional system. Let the reactants be confined to a one-dimensional 
box of length $L$ with reflecting boundary conditions at $x=0$ and $x=L$
and let $T(x)$ indicate the imposed temperature gradient across the box, 
or any arbitrary temperature profile with the same boundary conditions, $T(0)=T_1, T(L)=T_2$.  
We define the space- and time-dependent chemical molar fractions, $\mathcal{P}_\sigma(x,t)$, 
$\sigma=F,P,L$. The equations then read
\begin{eqnarray}
\label{e:reactdiff}
\frac{\partial \mathcal{P}_F(x,t)}{\partial t} &=& D \frac{\partial^2 \mathcal{P}_F(x,t)}{\partial x^2} 
                                       + k_{LF}(x) \mathcal{P}_{L}(x,t) - k_{FL}(x) \mathcal{P}_{F}(x,t) 
                                       \nonumber\\
\frac{\partial \mathcal{P}_P(x,t)}{\partial t} &=& D \frac{\partial^2 \mathcal{P}_P(x,t)}{\partial x^2} 
                                       + k_{LP}(x) \mathcal{P}_{L}(x,t) - k_{PL}(x) \mathcal{P}_{P}(x,t) 
                                       \\
\frac{\partial \mathcal{P}_L(x,t)}{\partial t} &=& D \frac{\partial^2 \mathcal{P}_L(x,t)}{\partial x^2} 
                                       + k_{FL}(x) \mathcal{P}_{F}(x,t) + k_{PL}(x) \mathcal{P}_{P}(x,t) + 
                                       \nonumber   \\
                                       && \qquad -(k_{LF}(x) + k_{LP}(x)) \mathcal{P}_{L}(x,t)\nonumber                                                                
\end{eqnarray}
where $D$ is the diffusion coefficient (assumed to be the same for all species) and
the rates are given by the obvious generalisations 
of the expressions introduced in the two-box model (see Fig.~\ref{f:3statesmodel}), that is,
\begin{equation}
\label{e:ratescont}
\begin{array}{lll}
k_{FL}(x) = \frac{\ds \mu}{\ds \eta_F} \,e^{-(E_L-E_F+\Delta E_F)/k_BT(x)} &\qquad
&k_{LF}(x) = \mu \,e^{-\Delta E_F/k_BT(x)} \\
k_{PL}(x) = \frac{\ds \mu}{\ds \eta_P} \,e^{-(E_L-E_P+\Delta E_P)/k_BT(x)} &\qquad
&k_{LP}(x) = \mu \,e^{-\Delta E_P/k_BT(x)}
\end{array}
\end{equation}
The steady-state populations, $\mathcal{P}_\sigma^\infty(x)=\lim_{t\to\infty} \mathcal{P}_\sigma(x,t)$,
are the solutions of the system obtained by letting the time derivatives vanish in Eqs.~\eqref{e:ratescont},
with the normalisation $\mathcal{P}_F^\infty(x) + \mathcal{P}_P^\infty(x) + \mathcal{P}_L^\infty(x) = 1$
(the stationary profiles are nearly flat). 
In the presence of a temperature gradient, a selection 
indicator can be defined to quantify the excess fraction of furanose, namely
\begin{equation}
\label{e:Rselcont}
\mathcal{R}_{\rm sel} = \frac{\ds \int_0^L \mathcal{P}_F^\infty(x)\,dx}
                             {\ds \int_0^L \mathcal{P}_P^\infty(x)\,dx}
                        \frac{\ds \int_0^L \mathcal{P}_P^{\rm eq}(x)\,dx}
                             {\ds \int_0^L \mathcal{P}_F^{\rm eq}(x)\,dx}     
\end{equation}
where $\mathcal{P}_\sigma^{\rm eq}(x)=\eta_\sigma 
\exp[-E_\sigma/k_BT(x)]/Z$ stands for the equilibrium distributions. By definition, 
for $T(x) = const.$, the selection indicator~\eqref{e:Rselcont} is unity. \\
\indent The plots reported in Fig.~\ref{f:Rsel} not only confirm the results obtained within the 
two-box model, but also reveal a remarkable fact. Nonequilibrium selection of furanose 
typically displays a maximum enhancement, corresponding to a restricted region of values of the temperature
gradient $\Delta T = T_2-T_1$ and energy barrier $\Delta E_F$. 
The maximum appears to correspond to a population boost of $40 - 50$ \%, for a wide
range of transport regimes (as gauged by the Damk\"ohler number) and 
kinetic parameters, notably the values of the barrier $\Delta E_P$.
It can be appreciated that in the infinite-transport  limit, Da $<1$, the optimum is 
progressively pushed to lower values of the energy barrier $\Delta E_F$.
Most intriguingly, the optimal 
value of temperature gradient appears to be consistently not far from the
gradients found in hydrothermal venting associated with extremely fresh lava flows
on the seafloor, with exit temperatures up to 407 $^{\rm o}$C recorded~\cite{Haase:2007aa}\,.
It appears therefore possible that optimal or suboptimal conditions could have  been met at certain spots
on the bottom of Hadean oceans.

%
\section{Summary and discussion}
%
%
\noindent In this paper we have delved into the question of why the $\beta-$furanose isoform of D-ribose
has been selected as the exclusive sugar component of nucleic acids instead of pyranose,
despite being far more unstable and hence present at lower molar fractions
in thermodynamic equilibrium up to relatively high temperatures. Our findings 
suggest that a plausible answer to this question is non-equilibrium enrichment 
beyond equilibrium sustained by a temperature gradient.\\
\indent In the first part of this work, we analyse the results of NMR
measurements of ribose isomerisation at  equilibrium at increasing temperatures. 
A simple thermodynamic equilibrium model reproduces the NMR data
to an excellent extent, revealing that the populations of the more unstable 
furanose species increase with temperature, while those of pyranoses decrease. 
As a result, a population inversion to a furanose-richer phase 
is predicted to occur at a temperature of about 150 $^{\rm o}$C, directly
connected to the large entropic degeneracy associated with furanose internal isoforms. 
Remarkably,  the same measurements repeated in the presence of a salt mixture simulating that of 
Hadean oceans led to an inversion temperature
about 20 $^{\rm o}$C lower. \\
\indent While high temperatures must have been ubiquitous in the early Earth  
environment, at least close to the rock/water interface,
it is highly unlikely that the kind of hotbeds suggested 
as possible primeval chemical reactors were at thermodynamic equilibrium.
Building on this idea, and on the physical parameters measured in our NMR experiments,
in the  second part of the paper we pursue the idea that furanose might have
been stabilised beyond the limits imposed by equilibrium thermodynamics under
non-equilibrium conditions. To this end, we investigate a simple model of ribose 
isomerisation in the presence of a steady temperature gradient. 
Our calculations show that driving the network far from equilibrium may 
lead to sustained {\em kinetic selection} of furanose, i.e. increased stabilisation with 
respect to thermal equilibrium, if linear-to-furanose interconversion proceeds appreciably
faster then linear-to-pyranose.\\
\indent The increased stabilisation occurs as diffusive and convective mass currents set in, 
shuttling molecules cyclically across the temperature gradient.
Thus, in the non-equilibrium steady state, sustained mass transport and chemical transformations 
may couple in such a way that furanose molecules absorb heat at the hot side and are transported 
to the cold end, where the extra heat is 
used to steer chemical relaxation towards the production of more furanose. 
We show that this scenario may emerge provided (i) the typical Damk\"ohler number of the system 
is lower than about $1-10$ (depending on energy barriers), 
and (ii) the relaxation of linear ribose to furanose is faster than its 
relaxation to pyranose. According to the first requirement, timescales characteristic 
of mass transport should roughly match those typical of chemical transformations. If this is the case, 
the fastest reactions will be kinetically selected, and the populations of 
associated products boosted beyond thermal equilibrium through sustained mass currents. 
The second requirement amounts to a condition on the relative magnitude of the energy 
barriers that separate chemical states in the system. While these are strictly irrelevant 
at equilibrium, in systems driven away from equilibrium, (kinetic) selection of the fittest may be
achieved as selection of the fastest~\cite{Busiello:2019aa,Zhang:2012aa}.
More generally, in nonequilibrium conditions, the fitness associated with 
a given energy landscape can be thought of as being {\em shifted} 
kinetically~\cite{Wang:2015aa,Wang:2008aa}.\\
\indent Variations of the temperature gradient and energy barriers have a profound effect 
on the extent of furanose nonequilibrium selection. We show that 
ultra-stabilisation is maximum in specific regions of parameter space. 
Intriguingly, optimal temperature gradients are predicted in the range $300-400$ $^{\rm o}$C, 
of the same order of magnitude as those found in the proximity of present-day 
venting activity at some spots on the seafloor~\cite{Haase:2007aa}\,. This prompts the 
intriguing hypothesis that {\em kinetic landscapes} such as those illustrated in Fig.~\ref{f:Rsel}
might have been integrated by evolution in more comprehensive, 
multi-fitness landscapes that eventually led to the biochemistry of life that we know today.\\
\indent In summary, an accurate thermodynamic characterisation of  ribose isomerisation
revealed the occurrence of population inversion between 
furanose and pyranose species at increasing temperatures in thermal equilibrium.
However, we demonstrated that the constraints imposed by equilibrium thermodynamics 
can be overcome and more furanose produced by driving the network far from equilibrium through 
a steady temperature gradient. \\
\indent It should be recognised that alternative possibilities that do not 
necessarily require a pre-existing abundance of ribofuranose for its selection 
have been proposed. In a more {\em systemic} perspective, these should be considered
at least as concurrent selection pathways.
Historically, the Formose reaction has been considered the major pathway, by
which ribose could be formed. However, detailed studies revealed the formation of
branched side products, low yields of ribose and high level of degradation that
paved the way for criticism~\cite{Larralde:1995aa}\,. The formation of
ribose bisphosphate from glycolaldehyde phosphate and formaldehyde, with similar
starting materials as the formose reaction, is suggested as an alternative
pathway by which ribose could have been driven to be a major 
product~\cite{Muller:1990aa}\,. But then again, the degradation kinetics of ribose and ribose
bisphosphate have always been of concern and one of the prime reasons for ribose
being dubbed a non-ideal candidate for the origin of RNA as genetic 
material~\cite{Larralde:1995aa}\,. It was suggested that ribose might have
been conserved in  contemporary metabolic pathways, such as the pentose
pathway (and subsequently into the glycolysis pathway), by kinetically
controlled reaction of ribose (eg. from formose reaction) with HCN, forming
stable lactone and aldonic acid which were later converted back to ribose after
the onset of enzymes~\cite{Larralde:1995aa}\,. More recently,
Eschenmoser suggested the necessity to revisit the HCN chemistry and suggested a
{\em Glyoxlate scenario} of primordial metabolism~\cite{Eschenmoser:2007aa}\,. Subsequent
studies on the reaction of dihydroxyfumarate (with glyceraldehyde) have shown
clean reactions where $\alpha$ and $\beta$ ribulofuranoses 
dominate~\cite{Sagi:2012aa}\,. Such a reaction could potentially integrate sugars into the modern
metabolic pathways.\\
\indent The abiotic origin of deoxyribonucleic acids is still
uncertain and it is unclear if the deoxynucleic acids emerged independently or
as a product of biotic selection. Our current study is limited to ribose and its
anomers. Nevertheless, within a non-equilibrium thermal gradient setting, a
similar argument in the case of the deoxyribofuranose might not be unreasonable,
given the fact that the 2-deoxyribose exhibits a similar anomeric distribution
as ribose, i.e. the deoxypyranoses dominate at 
equilibrium~\cite{Cortes:1991aa} up to 36 $^{\rm o}$C. 
Thus, we expect 2-deoxyribofuranoses could also be enriched using our model. 
Nonetheless, such a result would also only account for
one of the pathways by which $\beta$-deoxyfuranoses might have been utilised as a
component of nucleic acids. However, alternate routes to the synthesis of
2-deoxyribofuranosides using photoredox chemistry exist that do not necessitate
an enrichment mechanism~\cite{Xu:2019aa}\,. Recently, it has been experimentally
demonstrated that an aldol reaction between acetaldehyde and glyceraldehyde,
promoted by amino nitriles, leads to the formation of 2-deoxyribose~\cite{Steer:2017aa}
with $\simeq$ 5 \% yields. In such a scenario, where a concentration mechanism is
necessitated in order to accumulate molecules, our non-equilibrium model would
be advantageous.\\
\indent It should be stressed that our study does not necessarily 
preclude different means
by which $\beta$-ribofuranoses or even ribose might have been selected as the
principal component of  nucleic acids. Eschenmoser elegantly reasoned the
various possibilities and limitations of other aldoses and their corresponding
nucleic acids as prime candidates with potential to be genetic 
polymers~\cite{Eschenmoser:2011aa}\,. 
He suggested that alternative sugar modules such as
hexopyranoses could have competed with ribose, but might have been
left behind in the prebiotic competition due to their inherent inability to act
as informational carriers of genetic information (due to weak base pairing
properties). The base pairing properties of these alternate nucleic acids
further faded when the deoxyhexopyranoses were substituted as the sugar units.
The lack of functional prerequisite is not the only cause for evolutionary
rejection, but excessive stability of the base pairing in duplex state could
also lead to an evolutionary exclusion strategy. Such is the case of
pentopyranosyls nucleic acids, which form extremely stable base 
pairs~\cite{Eschenmoser:2011aa}\,. \\
\indent Our study in its present state of development is restricted to investigating
the influence of thermal equilibrium and non-equilibrium conditions 
on the relative anomeric ratios in a D-ribose enriched system. It cannot be
neglected that the Hadean Earth was far from such a model system. Factors like
pH, loss of  molecules due to dilution, presence of various minerals,
reactivity and degradation would affect this isomerisation phenomenon to varying
extents. 
Concerning pH or other non-thermal effects, it should be noted that,
unless pH (or other factors)  stabilizes ribofuranose {\em intrinsically}, 
a pH gradient would also be effective as long as the barriers of 
the transitions between the states are modulated by local pH, as they are by temperature. \\
\indent Sugars have been successfully synthesized under alkaline, mineral-assisted conditions.
Minerals like borates~\cite{Kim:2016aa} and 
silicates~\cite{Lambert:2004aa,Lambert:2010aa,Lambert:2014aa}  
are known to enhance the formation of pentoses. In fact, the most
efficient ways in which such minerals are able to stabilise these sugars are
with the furanoses that complex with these minerals due to the favourable
dihedral angle of their hydroxyl groups. It is also important to
note that the degradation pathways of the sugars are mainly 
via the open-chain~\cite{Furukawa:2017aa}
and the pyranose forms~\cite{Garrett:1969aa}\,,
which are more reactive in comparison to the furanoses, thus supporting the fact that the
ribofuranoses could also serve as a shielding mechanism against the degradation of
sugars and in turn enhance their concentration and promote complexation with
minerals like borates and silicates (known to be abundant on the early Earth).
With respect to the reactivity of these sugars, it has been demonstrated that a
borate-complexed ribose is regioselectively phosphorylated under dry 
conditions~\cite{Kim:2016aa} and in a biphasic system of aqueous 
formamide~\cite{Furukawa:2015aa}\,. 
More recently, under non-equilibrium thermal conditions, phosphorylation
of nucleoside has been demonstrated successfully~\cite{Morasch:2019aa}\,.\\
\indent In summary, our study should be considered as an attempt to 
quantify one among several possibilities through which the
emergence of contemporary nucleic acid component $\beta$-ribofuranose
would be favoured. In this regard, our results demonstrate the subtle  
non-equilibrium physicochemical effects that may arise from the interplay 
of vast chemical landscapes and the geophysical conditions of their surroundings 
on primordial Earth.

\section*{Acknowledgements}

AVD and FW would like to acknowledge Region Centre, France, for a doctoral bursary for AVD.
PDLR and SL thank the Swiss National Science Foundation for support under grant 200020\_178763.

%
%
%
%
\providecommand{\latin}[1]{#1}
\makeatletter
\providecommand{\doi}
  {\begingroup\let\do\@makeother\dospecials
  \catcode`\{=1 \catcode`\}=2 \doi@aux}
\providecommand{\doi@aux}[1]{\endgroup\texttt{#1}}
\makeatother
\providecommand*\mcitethebibliography{\thebibliography}
\csname @ifundefined\endcsname{endmcitethebibliography}
  {\let\endmcitethebibliography\endthebibliography}{}

%
%

\newpage
\centerline{\bf\Huge Supplementary material}

\section{NMR experiments}

\subsection{Choice of the relaxation time $T_1$ in the NMR experiments}

\noindent The first step in order to quantify the relative 
population of each anomer through $^{13}$C NMR experiments is to evaluate 
the relaxation time $T_1$ for all species. Since $T_1$ is strongly dependent on temperature, 
$T_1$ measurements were carried out at each temperature before the spectral data were recorded. 
The results of these experiments are reported in Table~\ref{tbl:T1}. 
%
%
\begin{table}[h]
\small
  \caption{\ Values of $T_1$ (seconds) for C1 of ribose at thermal equilibrium 
             at different temperatures.}
  \label{tbl:T1}
  \begin{tabular*}{0.48\textwidth}{@{\extracolsep{\fill}}lllll}
    \hline
    Temperature ($^{\rm o}$C) & $\alpha$P & $\alpha$F &  $\beta$P & $\beta$F \\
    \hline
    20  &   1.99	   &  1.59	   &  1.79	   &  1.87  \\
    30  &   2.37	   &  2.85	   &  2.42	   &  2.15  \\
    40  &   2.74	   &  3.62	   &  2.77	   &  2.91  \\
    50  &   3.77	   &  3.56	   &  3.51	   &  3.74  \\
    60  &   5.06	   &  5.49	   &  5.19	   &  5.10  \\
    70  &   5.79	   &  6.44	   &  5.11	   &  5.18  \\
    80  &   7.94	   &  5.16	   &  6.11	   &  5.79  \\
   \hline
  \end{tabular*}
\end{table}
%
It can be appreciated that $T_1$ increases significantly with temperature from about 2 to 5 seconds, 
for temperatures between 10 $^{\rm o}$C and 80 $^{\rm o}$C, the temperature range of this study. 
Overall, however, it is fair to say that the values of $T_1$ are similar for all anomers at 
a given temperature. Only  $\alpha$pyranose shows a significantly higher value at the higher 
temperature of 80 $^{\rm o}$C.Building on these results, we investigated 
the anomerisation of ribose  by $^{13}$C NMR at different temperatures, 
by adapting the $D_1$ ($5 T_1$) for each experiment.

\subsection{Extracting the equilibrium molar fractions from the NMR spectra}

\begin{figure}[t!]
\centering
\includegraphics [width=14truecm]{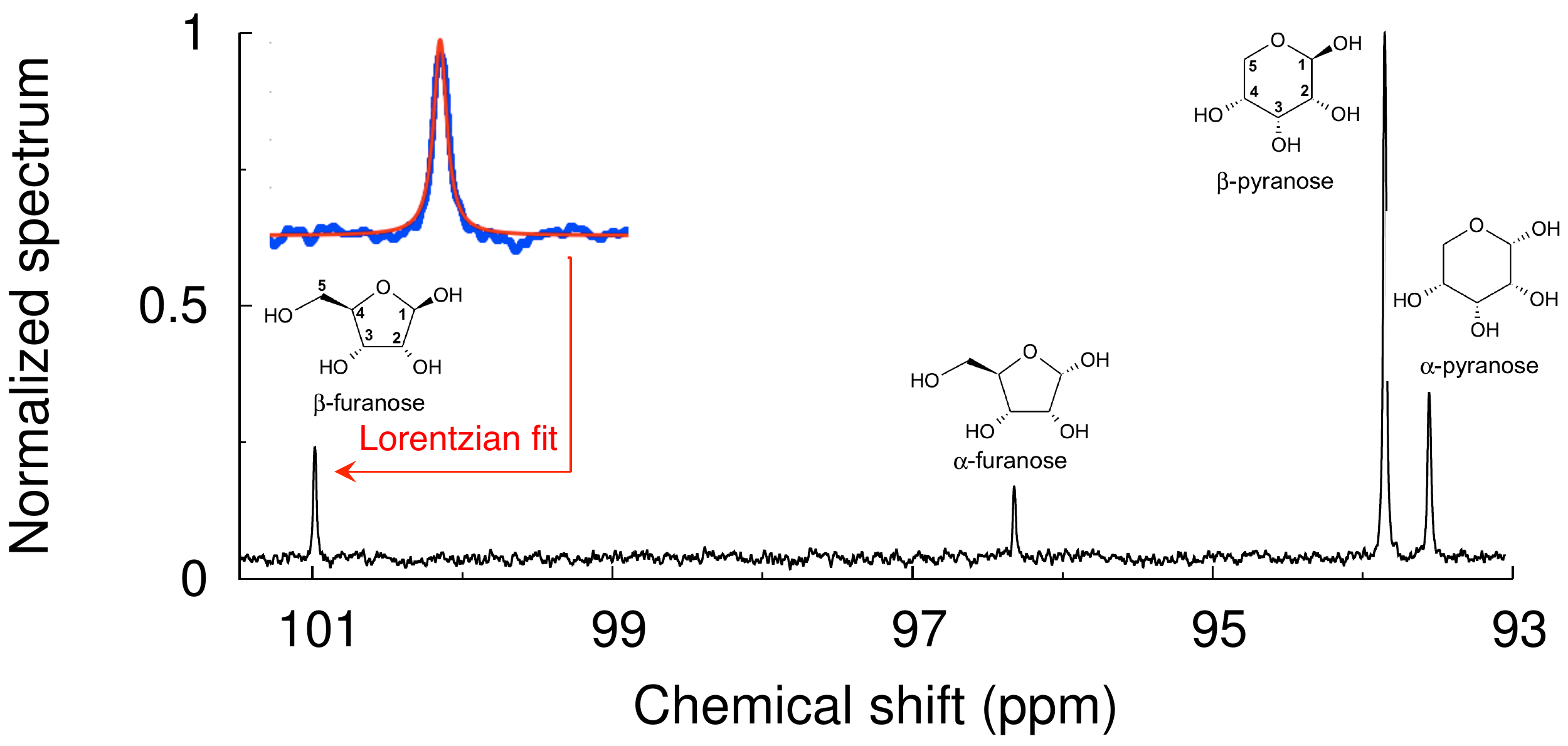}
\caption{$^{13}$C NMR spectrum of D-ribose in solution at $T=25$ $^{\rm o}$C and ambient pressure.  
The relative fractions of the four species 
$x_i$ were estimated by fitting the areas $A_i$ below the 
respective NMR peaks, i.e. $x_i = A_i/\sum_m A_m$.
The inset  shows a typical Lorentzian fit of one of the anomers' peak.}
\label{f:NMRsp}
\end{figure}

\noindent At 25 $^{\rm o}$C ($T_1=1.8$ s), 
the signals of C1-$\alpha$P,  C1-$\beta$P,  C1-$\alpha$F and  C1-$\beta$F are found, 
respectively, at 93.56 ppm, 93.85 ppm 96.33 ppm and 100.99 ppm. 
A large number of $^{13}$C NMR spectra, between 50 and 80, was recorded for each temperature 
and the  average relative molar fractions computed by fitting the corresponding peak areas with 
Lorentzian line shapes. 
This procedure is illustrated for one representative spectrum in Fig.~\ref{f:NMRsp}.

%

\subsection{Model Hadean water}

\noindent The composition of the artificial sea water~\cite{Dass:2018aa} used
in our NMR experiments is reported in Table~\ref{tbl:ocean}.

%
\begin{table}[h!]
\small
  \caption{\ Salt composition of the simulated Hadean sea water used in our $^{13}$C NMR 
  experiments, pH 6.3.}
  \label{tbl:ocean}
  \begin{tabular*}{0.48\textwidth}{@{\extracolsep{\fill}}lr}
    \hline
    Compound & g/L \\
    \hline
    NaCl	                &   70.00   \\
    NaSiO$_2$           &   	 0.30  \\
    FeCl$_2$            &   	 0.30  \\
    KCl	                &    1.40   \\
    NaHCO$_3$           &   	 0.10  \\
    KBr	                &    0.30   \\
    H$_3$BO$_3$	        &    0.20   \\
    NaF	                &    0.50   \\
    MgCl$_2$. 6H$_2$O	&   19.30   \\
   \hline
  \end{tabular*}
\end{table}
%
%
\begin{itemize}
\item NaCl in the modern ocean is 35g/L, and has maximum values at the vents of 
hydrothermal systems. Fluid inclusion study demonstrates that the concentration of 
NaCl exceeds current levels in both Archaean seawater and hydrothermal fluid. 
Zharkov~\cite{Zharkov:1981aa} supported an Archaean sea of 1.2 times the current salinity, 
inferring lacking complexity of biological evolution. This was refuted by Knauth~\cite{Knauth:1998aa}, 
who advocated an ocean at least 1.6 times as saline as the 
modern. de Ronde et al.~\cite{De-Ronde:1997aa} conducted detailed fluid inclusion studies in 
Barberton Greenstone Belt ironstones to assign ionic concentrations to the water in which those formed. 
We have selected the hydrothermal end-member of their dataset as being the example which most closely 
represents the environment we aim to simulate. Therefore, the Archaean ocean at the sediment-hydrothermal 
water interface is approximately twice the salinity of the modern ocean, 70 g/L. Rollinson~\cite{Rollinson:2007aa} 
suggests that the Archaean ocean was saturated in both the ions of Na and Cl.

\item NaSiO$_2$ is added to provide the Si ion. The continued addition of Na is inconsequential, 
as explained above it was oversaturated. Knauth~\cite{Knauth:2005aa} suggests that in an ocean of 50 $^{\rm o}$C, 
Si would have a concentration of 300ppm, however, Rimstidt~\cite{Rimstidt:1997aa} had previously given a conservative 
estimate of only 20ppm. Early Archaean stratigraphy is profoundly and ubiquitously silicified throughout, 
particularly in the hydrothermally influenced shallow water environments which we aim to simulate in this experiment. 
Therefore, we chose the value of 300ppm (0.3 g/L), with the observation that this is perhaps even a conservative estimate 
of the waters. Knauth’s estimation was for the whole ocean; unfortunately, the modern hydrothermal analogues are 
not strictly equivalent to ancient examples.

\item FeCl$_2$ was chosen to replace the FeSO$_4$ of the modern ocean, and is specifically to add Fe; 
as mentioned above, the saturation of the ocean in Cl~\cite{Rollinson:2007aa}  means that its overaddition 
should not affect our estimation. SO$_4$ was low to negligible in the Hadean-Archaean 
ocean~\cite{Habicht:2002aa,Sumner:1997aa}, but there was a substantial Fe dissolved 
reservoir~\cite{Crowe:2008aa,Holland:1986aa}; Fe in Archaean carbonate indicates that concentrations in 
the equilibrating ocean were twice their current values. We consider the example of Lac Pavin, France, a 
supposed Archaean analogue water body featuring an iron system~\cite{Busigny:2014aa}. Throughout the lake profile, 
Fe concentrations vary from 2 $\mu$M in the top waters to 2,400 $\mu$M in the iron-rich waters at the base; we select 
the latter value as representative of the hydrothermal scenario that we propose. This value is similar to those of 
Archaean and modern hydrothermal zone Fe concentration described 
by Douville et al. ~\cite{Douville:2002aa,Hawkes:2013aa,Gallant:2006aa}.
It exceeds the 40-120 $\mu$M suggested by Canfield~\cite{Canfield:2005aa,Kappler:2005aa,Crowe:2008aa},
though their estimations represent oceanic concentrations, which we consider to be the seawater end-member. 

\item KCl is added to provide the K ion, which would have been sourced from effusions of hydrothermal 
fluids and dissolution of feldspar. Rollinson~\cite{Rollinson:2007aa}  suggests that K concentration was 19 mM 
in the Archaean ocean. We consider that this estimate may be conservative, however, there would have been a limited 
flux from erosion of the seafloor which, being komatiite-basalt in composition, has low, but not negligible, K content.

\item NaHCO$_3$ provides the bicarbonate, and Na, which was saturated. Using the relationships of dissolved 
carbonates in the ocean, we conclude that at pH 6, 70 \% (0.7 mole fraction) of total carbonate would have been 
bicarbonate. Thus, of the three phases in the carbonate diagram, bicarbonate would be the dominant phase. 
The presence of carbonates in the Archaean is understood only poorly due to a paucity in their preservation. 
We have used the pre-industrial value of bicarbonate, 1, 757 $\mu$M/L, to ensure that the ion is present, 
though this is the least well-constrained component.

\item KBr is added for the Br ion which, from fluid inclusion studies of 3.23 Ga ironstone pods in Barberton, 
is estimated at 2.59 mM/L for the hydrothermal end-member. A corollary of this addition is the addition of K, 
which has already been accounted for in KBr.

\item H$_3$BO$_3$ is chosen for the BO$_3$ ion. Modern values are 4-5 ppm (0.005 g/L), which we 
take as the absolute minimum for Archaean values. Since borate is associated with hydrothermal activity,
we have taken the values of current hydrothermal fluids at 203 $^{\rm o}$C 
as indicative of the levels of borate present in our simulated hydrothermal scenario. 
Therefore, 0.2 g/kg of the compound is needed~\cite{Ricardo:2004aa}. 

\item NaF is added for the F ion, which is major in hydrothermal effusions. No data for F was extracted 
from the ironstone fluid inclusions of de Ronde et al.~\cite{De-Ronde:1997aa}, no from any other example in 
our extensive literature survey. We thus turned to modern submarine hydrothermal systems as the closest 
available analogue. Values include 44 $\mu$M/L, 36.2-65.4 $\mu$M (of which the 40.2-42.9 $\mu$M bracket (cool flange) 
may be most representative: island arc hydrothermal field, and 500 ppm in an ore-forming fluid 
(since this fluid is an ore-forming example, the value is likely a slight overestimation of our more 
passive hydrothermal scenario).

\item MgCl$_2$.6H$_2$O is added for Mg, and since the ultramafic oceanic crust was rich in 
olivine-containing rocks (komatiite and tholeiite), we have taken the upper limit estimated by 
Rollinson~\cite{Rollinson:2007aa} of 95 mM/L.

\item CaCl$_2$.2H$_2$O is added for Ca, and since pyroxene was a major mineral in the 
reactive (dissolution of volcanic glass and feldspar) ultramafic oceanic crust, we 
have taken the upper limit estimate by Rollinson~\cite{Rollinson:2007aa} of 50 mM/L.

\end{itemize}

\newpage
\centerline{\Huge \bf Theory}
\bigskip

\noindent In the following, the energy of furanose  (F) is referred to as $E_1$,  
that of pyranose (P) as $E_2$, while that of the transition state, linear sugar (L) is 
referred to as $E_0$.
 
\section{Diagrammatic calculation of the steady-state molar fractions}

\noindent With reference to the node numbering shown in Fig.~\ref{f:3statesmodel},
the rate equations associated with the reduced reaction network (one high-energy (F) state $E_1$, 
one low-energy (P) state $E_2$ and a transition state, $E_0$) read
\begin{eqnarray}
\label{e:rateeqn}
&&\dot{x}_0 = k_{10} x_1 + k_{50} x_5 + k_D x_3 - \left( k_{01} + k_{05} + k_D \right) x_0  \nonumber\\
&&\dot{x}_1 = k_{01} x_0 + k_Dx_2 - \left( k_{10} + k_D\right) x_1                          \nonumber\\
&&\dot{x}_2 = k_{32} x_3 + k_Dx_1 - \left( k_{23} + k_D\right) x_2                          \\
&&\dot{x}_3 = k_{23} x_2 + k_{43} x_4 + k_D x_0 - \left( k_{32} + k_{34} + k_D \right) x_3  \nonumber\\    
&&\dot{x}_4 = k_{34} x_3 + k_D x_5 - \left( k_{43} + k_D\right) x_4                         \nonumber\\
&&\dot{x}_5 = k_{05} x_0 + k_D x_4 - \left( k_{50} + k_D\right) x_5                         \nonumber                                     \end{eqnarray}
%
%
%
\begin{figure}[h!]
\centering
\includegraphics[width=11truecm]{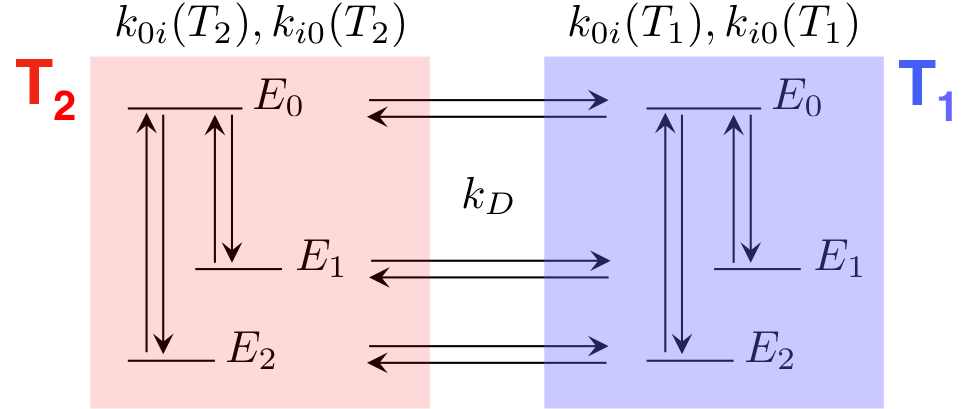}
\includegraphics[width=7truecm]{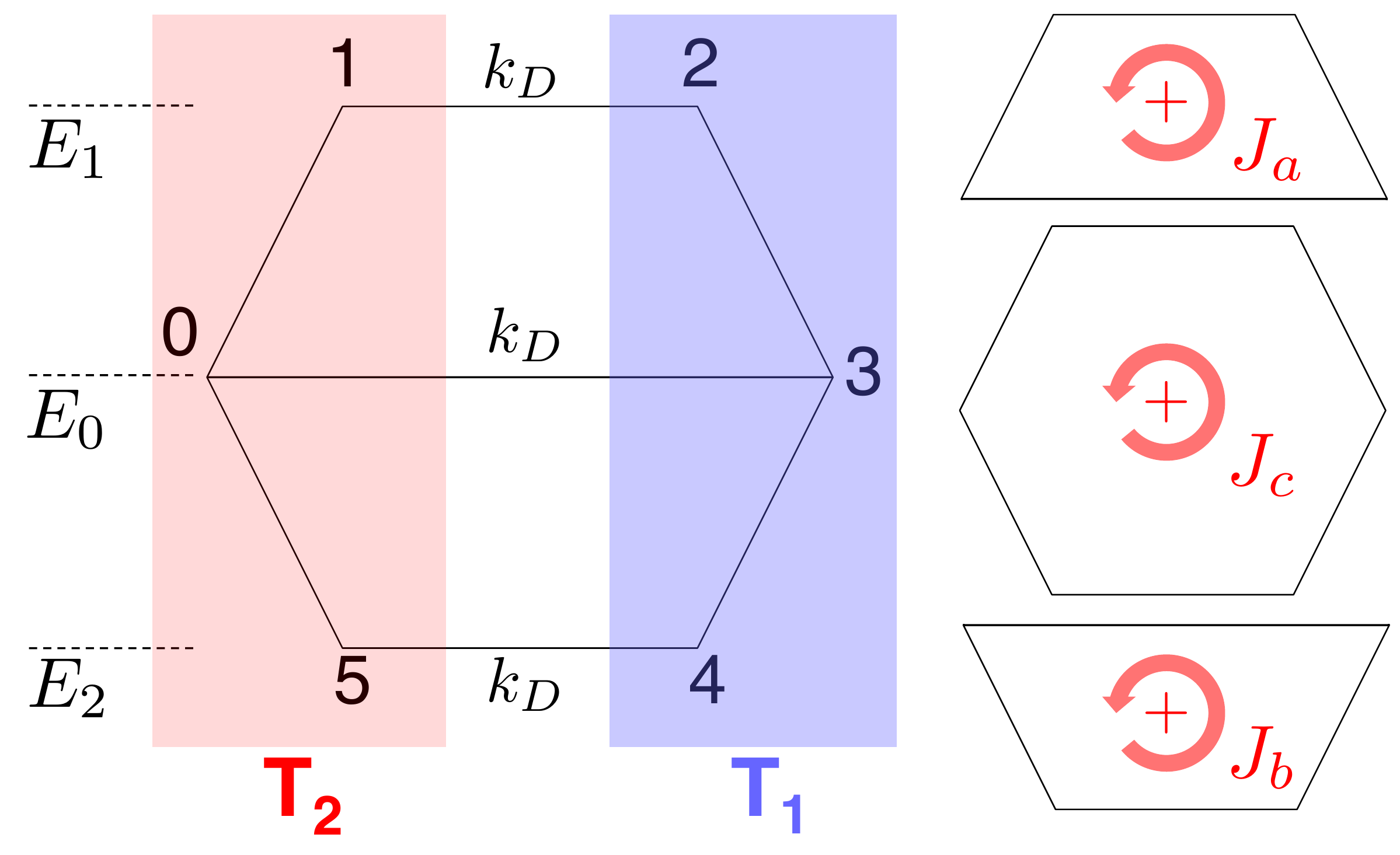}
\includegraphics[width=6truecm]{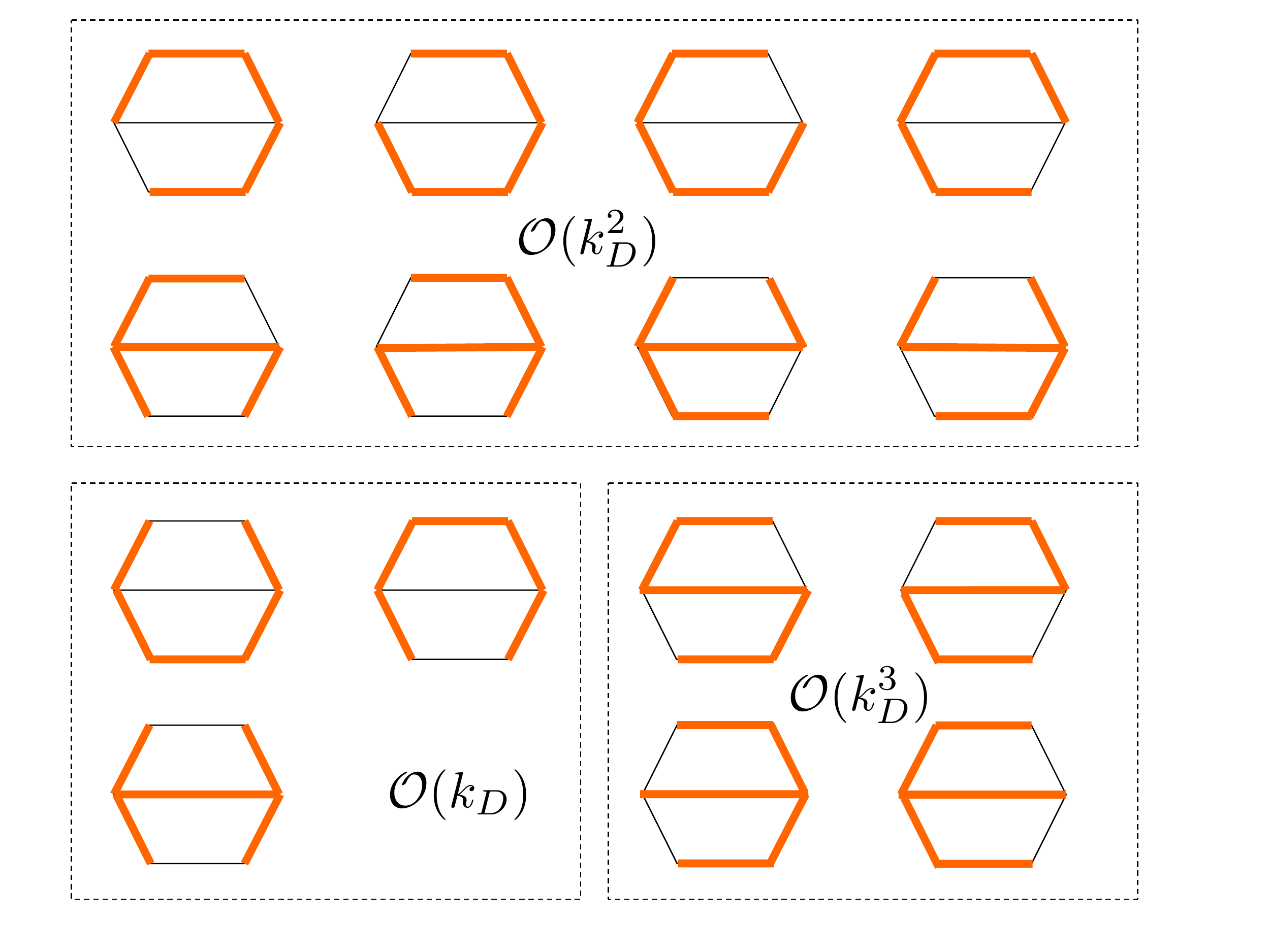}
\caption{(Top) reduced model of ribose isomerisation reactions in a steady temperature 
gradient ($T_2 > T_1$). The highest-energy linear species has energy $E_0$. The two  
furanose species have been coalesced into the high-energy species 1 (energy $E_1$),
while the two pyranose species represent the ground state in this model. Each species 
can diffuse or be advected across the temperature gradient with a global rate $k_D$.
Bottom left: Graph corresponding to the above chemical network with identification of the nodes
and illustration of the three cycle fluxes present in this model. The convention for positive 
cycles is shown explicitly. Bottom right: the complete set of partial diagrams, each of which 
contains the maximum number of lines (five here) that can be included without forming a cycle.
The partial diagrams are grouped according to the number of {\em mobility} links that they 
contain (i.e. branches giving  a contribution proportional to $k_D$).}
\label{f:3statesmodel} 
\end{figure}
%
%
with the normalisation $\sum_{m=0}^5x_m=1$. The explicit expressions for the rates read (see also 
main text):
\begin{equation}
\label{e:rates}
\begin{array}{lcl}
k_{01} = \mu \,e^{-\beta_2\Delta E_1} &\quad& k_{32} = \mu \,e^{-\beta_1\Delta E_1} \\
k_{05} = \mu \,e^{-\beta_2\Delta E_2} &\quad& k_{34} = \mu \,e^{-\beta_1\Delta E_2} \\
k_{10} = \frac{\ds \mu}{\ds \eta_1} \,e^{-\beta_2(E_0-E_1+\Delta E_1)} &\quad& 
k_{23} = \frac{\ds \mu}{\ds \eta_1} \,e^{-\beta_1(E_0-E_1+\Delta E_1)} \\
k_{50} = \frac{\ds \mu}{\ds \eta_2} \,e^{-\beta_2(E_0-E_2+\Delta E_2)} &\quad& 
k_{43} = \frac{\ds \mu}{\ds \eta_2} \,e^{-\beta_1(E_0-E_2+\Delta E_2)}
\end{array}
\end{equation}
where $\beta_i^{-1} = k_BT_i$, $i=1,2$.
Following Ref.~\citenum{Hill:1989aa}, the stationary state of the rate equations~\eqref{e:rateeqn}
can be computed by referring to the complete set of {\em partial diagrams} of the 
network (see Fig.~\ref{f:3statesmodel}). The steady-state value of state $m$ is 
then obtained by introducing a directionality to each line in each partial diagram,
so that all connected path {\em flow} toward the vertex $m$. Each directional line 
corresponds to a rate constant, hence each {\em directional} partial diagram corresponds
to the product of five rate constants. The prescription is then that each $x_m$ is 
proportional to the sum of all the associated directional diagrams. 
It will be remarked that the partial diagrams contain either 1,2 or 3 {\em mobility}
link, and therefore they can be classified as corresponding to terms 
of order $k_D^n$ ($n=1,2,3$ in the expressions for the stationary probabilities (molar 
fractions). 
%
\begin{figure}[t!]
\centering
\includegraphics[width=8truecm]{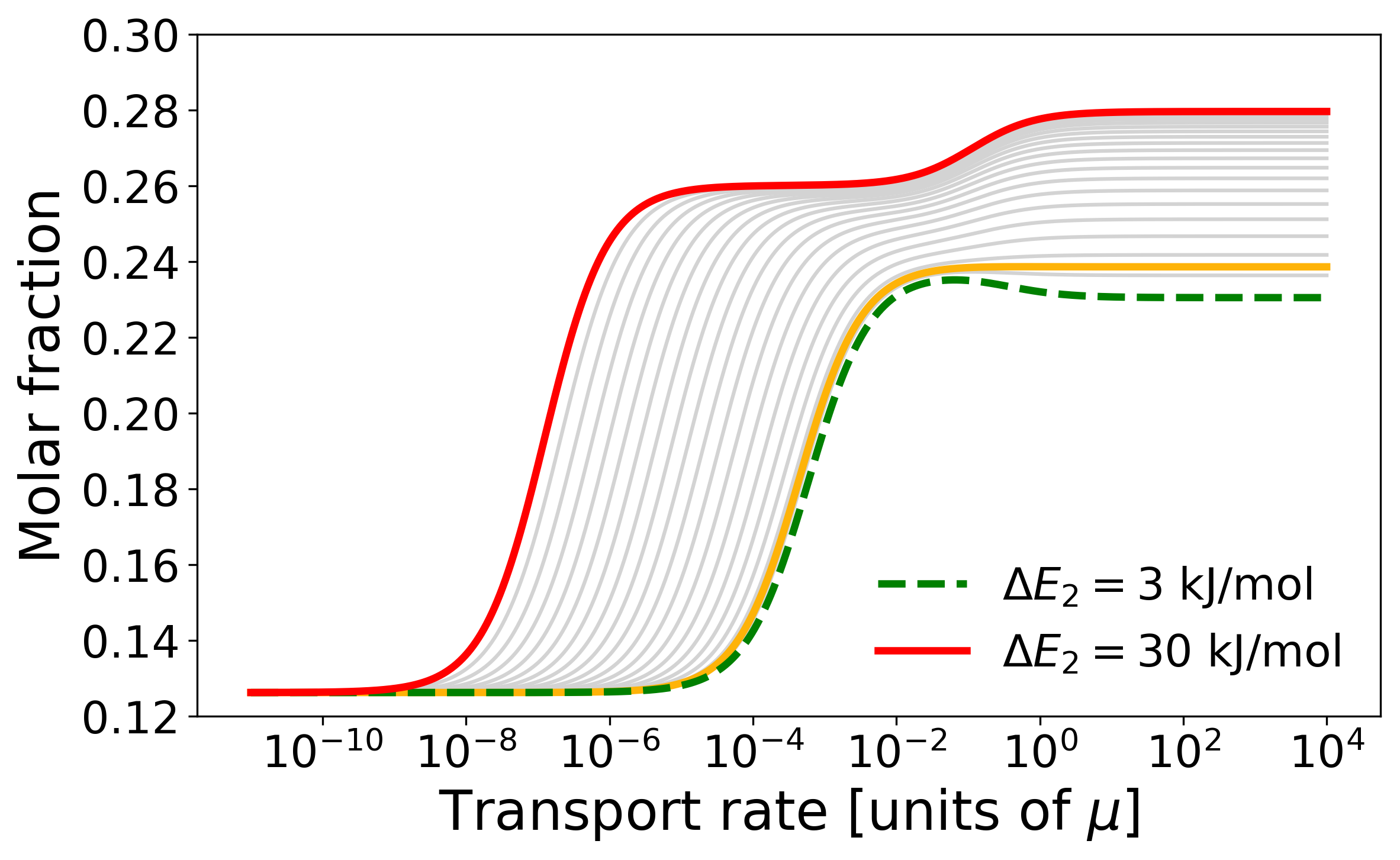}
\includegraphics[width=8truecm]{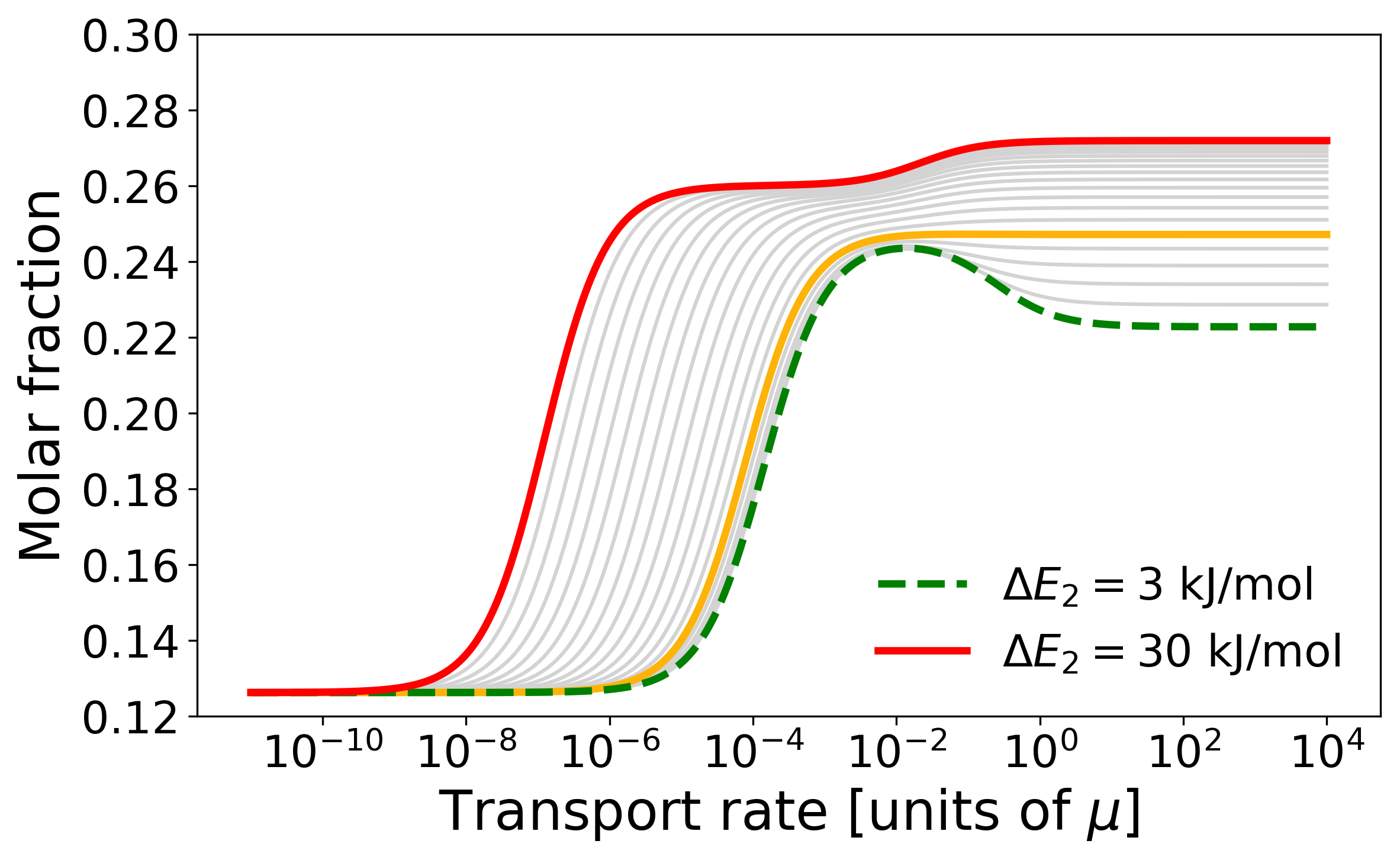}
\includegraphics[width=8truecm]{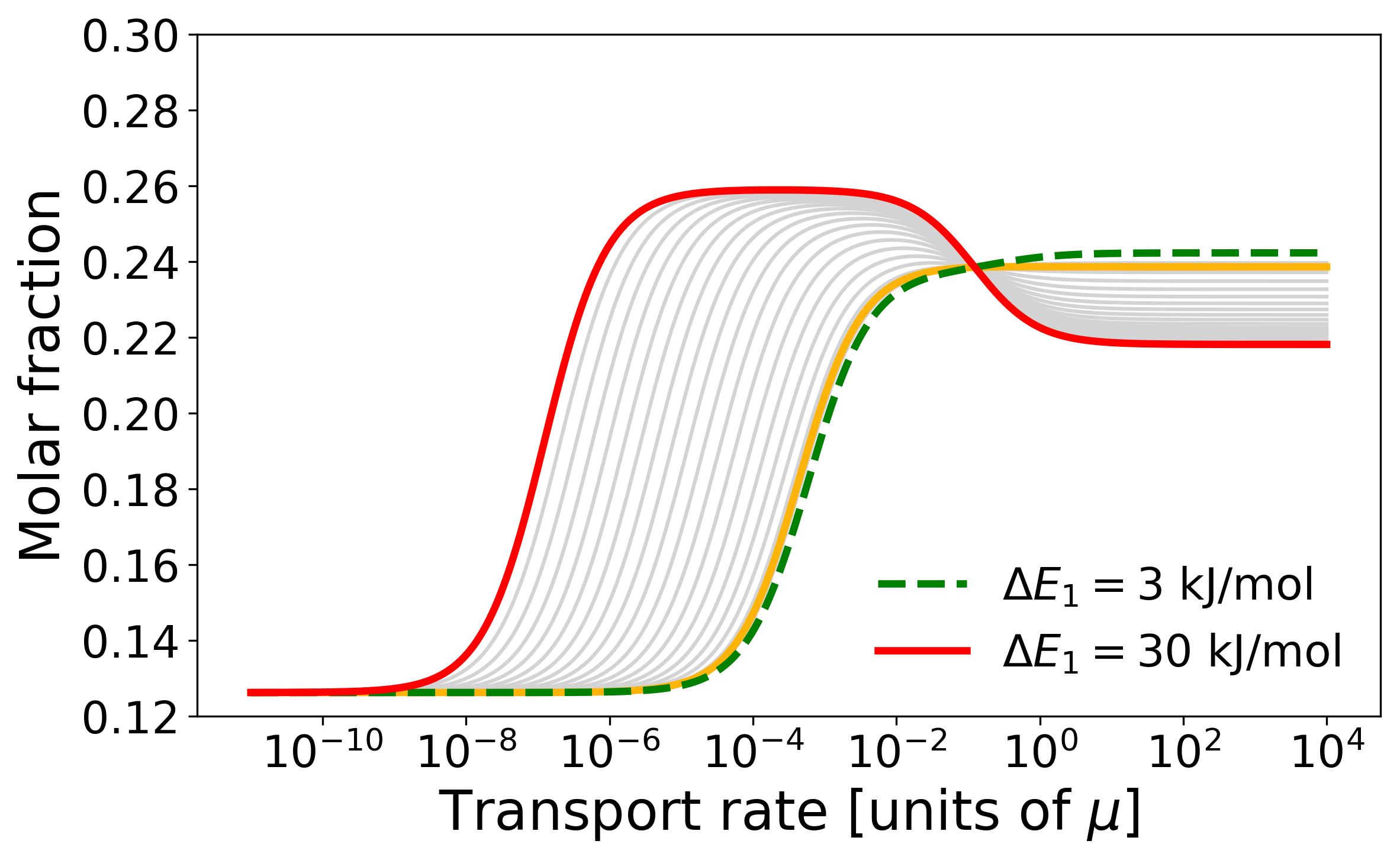}
\includegraphics[width=8truecm]{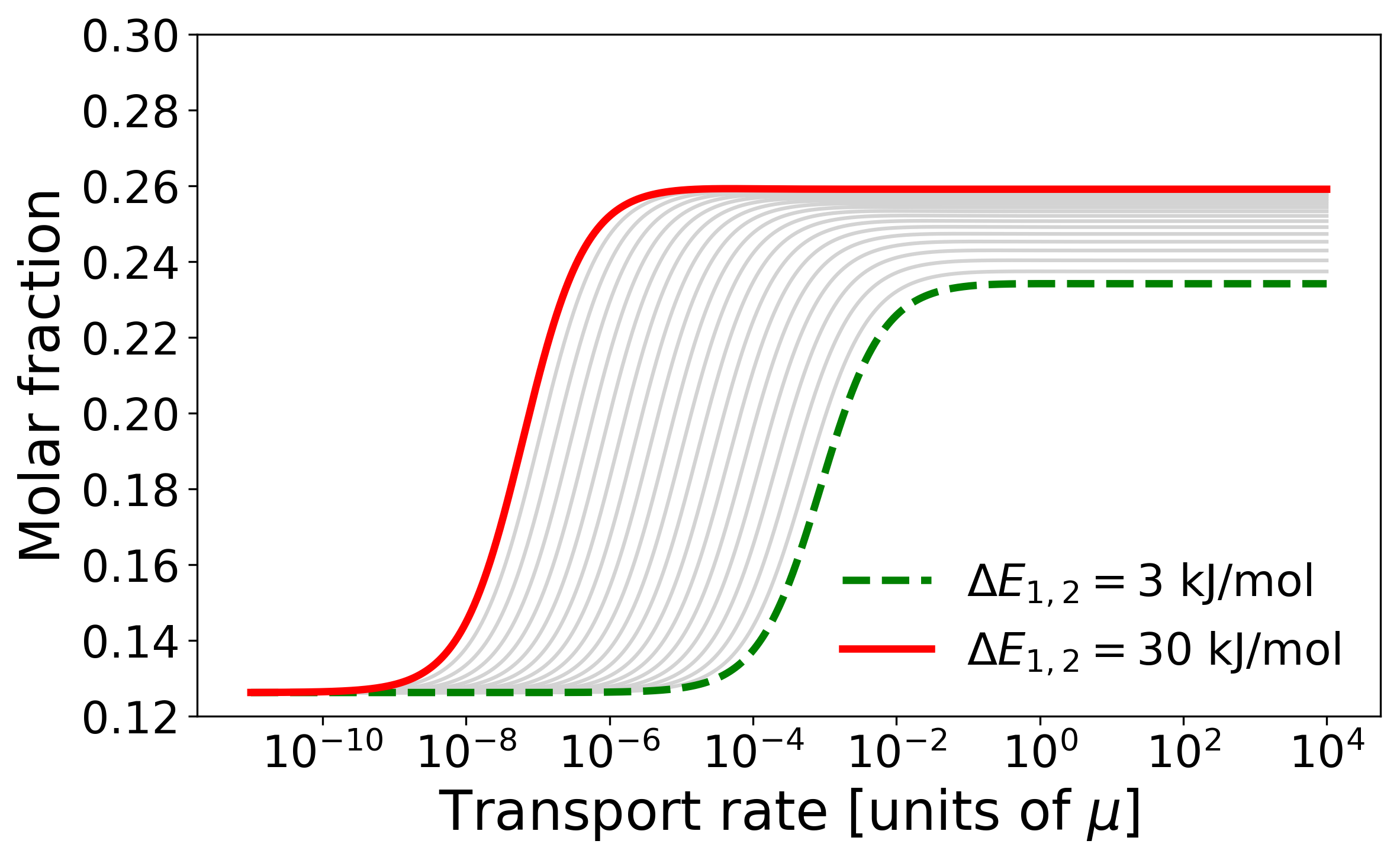}
\caption{\textbf{Illustration of the ultra-stabilisation/ultra-destabilisation effects}.
Molar fraction of the high-energy furanose species in a steady 
thermal gradient of the same order as that present between the top 
and the base of a thermal vent ($T_2=210\,^{\rm \,o}$C, $T_1=60\,^{\rm \,o}$C) as a function
the transport rate $k_D$ (in units of $\mu = k_{01}^\infty = k_{02}^\infty$)
for different choices of the energy barriers $\Delta E_1, \Delta E_2$. 
Clockwise from top to bottom: 
$\Delta E_1 = 5$ kJ/mol and varying $\Delta E_2 \in [3,30]$ kJ/mol,
$\Delta E_1 = 10$ kJ/mol and varying $\Delta E_2 \in [3,30]$ kJ/mol,
$\Delta E_2 = 5$ kJ/mol and varying $\Delta E_1 \in [3,30]$ kJ/mol
(the amber lines correspond to $\Delta E_2= \Delta E_1$),
varying $\Delta E_1 = \Delta E_2 \in [3,30]$ kJ/mol.
Parameters are $\eta_1 = 29.34$, $\eta_2 = 1.945$, $E_0=19$ kJ/mol, $E_1=13.6$ kJ/mol, 
$E_2=3.1$ kJ/mol, corresponding to the average values for the two furanose and pyranose
enantiomers measured from equilibrium NMR experiments (see Table~I in the main text).}
\label{f:xiD1D20-30} 
\end{figure}
%
%
%
%
As an example, considering state $1$, the steady-state probability reads
\begin{equation}
\label{e:ss1}
x_1 = \frac{k_D^3(k_{43} k_{01} + k_{50}k_{32} ) + 
            k_D^2(k_{43}k_{32}k_{01} + [\dots]_7) + 
            k_D(k_{50}k_{01}k_{23}k_{34} + [\dots]_2)}
            {k_D^3(k_{43} k_{01} + k_{50}k_{32} ) + 
            k_D^2(k_{43}k_{32}k_{01} + [\dots]_7) + 
            k_D(k_{50}k_{01}k_{23}k_{34} + [\dots]_2) + [\dots]_{65}}
\end{equation}
where the notation $[\dots]_n$ stands for $n$ {\em other} terms.
Overall, each expression such as eq.~\eqref{e:ss1} contains 13 terms 
in the numerator (i.e. as many as there are partial diagrams) and 78 terms 
in the denominator (i.e. 13 terms multiplied 6 nodes). As stated in the 
main text, such expressions can be cast in the general form of ratios 
of second-order polynomials, i.e.
\begin{equation}
\label{e:pssgen}
x_i = \frac{\alpha_{2i} k_D^2 + \alpha_{1i} k_D + \alpha_{0i}}
           {A_{2} k_D^2 + A_{1} k_D + A_0}, \qquad i=0,1,\dots,5
\end{equation}
with 
\begin{equation}
\label{e:pssgenN}
A_n =\sum_{j=0}^5\alpha_{nj} , \qquad n=0,1,3
\end{equation}
%
\subsection{The limit $k_D\to 0$, thermal equilibrium}

\noindent In our model, the thermal equilibrium scenario corresponds to two  
isolated boxes, each reaching thermal equilibrium at a separate temperature. 
Mathematically, this corresponds to the limit $k_D\to0$, Da$\,\propto k_D^{-1}\to \infty$,
the steady-state populations reducing to their respective equilibrium expressions 
(see Eqs. (2) in the main text),
\begin{equation}
\label{e:xeqkD0}
x_i^{\rm eq} \stackrel{\rm def}{=} \lim_{k_D\to 0} x_i = 
\frac{\alpha_{0i}}{A_0} 
\end{equation}
More precisely, we have 
\begin{equation}
\label{e:xieqexpl}
\begin{array}{ll}
x_1^{\rm eq} = \frac{1}{2} P_{E_1}^{\rm eq}(T_2) & 
x_2^{\rm eq} = \frac{1}{2} P_{E_1}^{\rm eq}(T_1) \\
x_5^{\rm eq} = \frac{1}{2} P_{E_2}^{\rm eq}(T_2) & 
x_4^{\rm eq} = \frac{1}{2} P_{E_2}^{\rm eq}(T_1) \\ 
x_0^{\rm eq} = \frac{1}{2} P_{E_0}^{\rm eq}(T_2) & 
x_3^{\rm eq} = \frac{1}{2} P_{E_0}^{\rm eq}(T_1) 
\end{array}
\end{equation}
with 
\begin{equation}
\label{e:Peq}
P_{E_i}^{\rm eq}(T) = \frac{\eta_i e^{-E_i/k_BT}}{Z(T)},
\end{equation}
where $Z(T) = e^{-E_0/k_BT} + \eta_1 e^{-E_1/k_BT} + \eta_2 e^{-E_2/k_BT}$
denotes the partition function at temperature~$T$. 
The factor $1/2$  in Eqs.~\eqref{e:xieqexpl} reflects the fact that 
each box contains half of the total mass.
%
%
\begin{figure}[t!]
\centering
\includegraphics [width=12.5truecm]{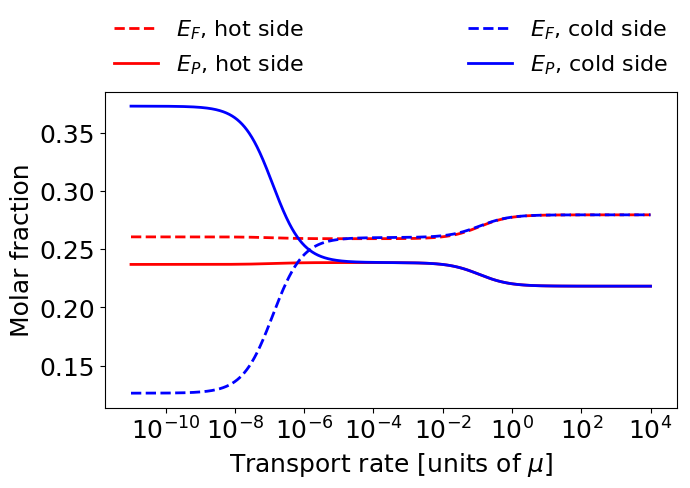}
\caption{{\bf The high-energy furanose species can be ultra-stabilised by sustained mass 
currents at  large enough transport rates}.
Molar fraction of furanose (dashed lines) and pyranose (solid lines) species in a steady 
thermal gradient with $T_2=210^{\rm \,o}$C, $T_1=60^{\rm \,o}$C vs
transport rate $k_D$ in units of $\mu$ (Da$^{-1}$) for 
$\Delta E_1 = 5$ kJ/mol, $\Delta E_2 = 30$ kJ/mol. 
Two consecutive stabilisation crossovers are observed as the transport rate increases, 
the final {\em ultra-stabilisation} transition occurring when Da $\simeq 1$ ($k_D \simeq \mu)$. 
Equilibrium parameters are $\eta_1 = 29.34$, $\eta_2 = 1.945$, $E_0=19$ kJ/mol, $E_1=13.6$ kJ/mol, 
$E_2=3.1$ kJ/mol, set as the average values for the two furanose and pyranose
enantiomers measured from equilibrium NMR experiments (see Table~I in the main text).}
\label{f:ultrastab} 
\end{figure}
%

\subsection{Finite values of the transport rate $k_D$, non-equilibrium steady states}

\begin{figure}[t!]
\centering
\includegraphics [width=14truecm]{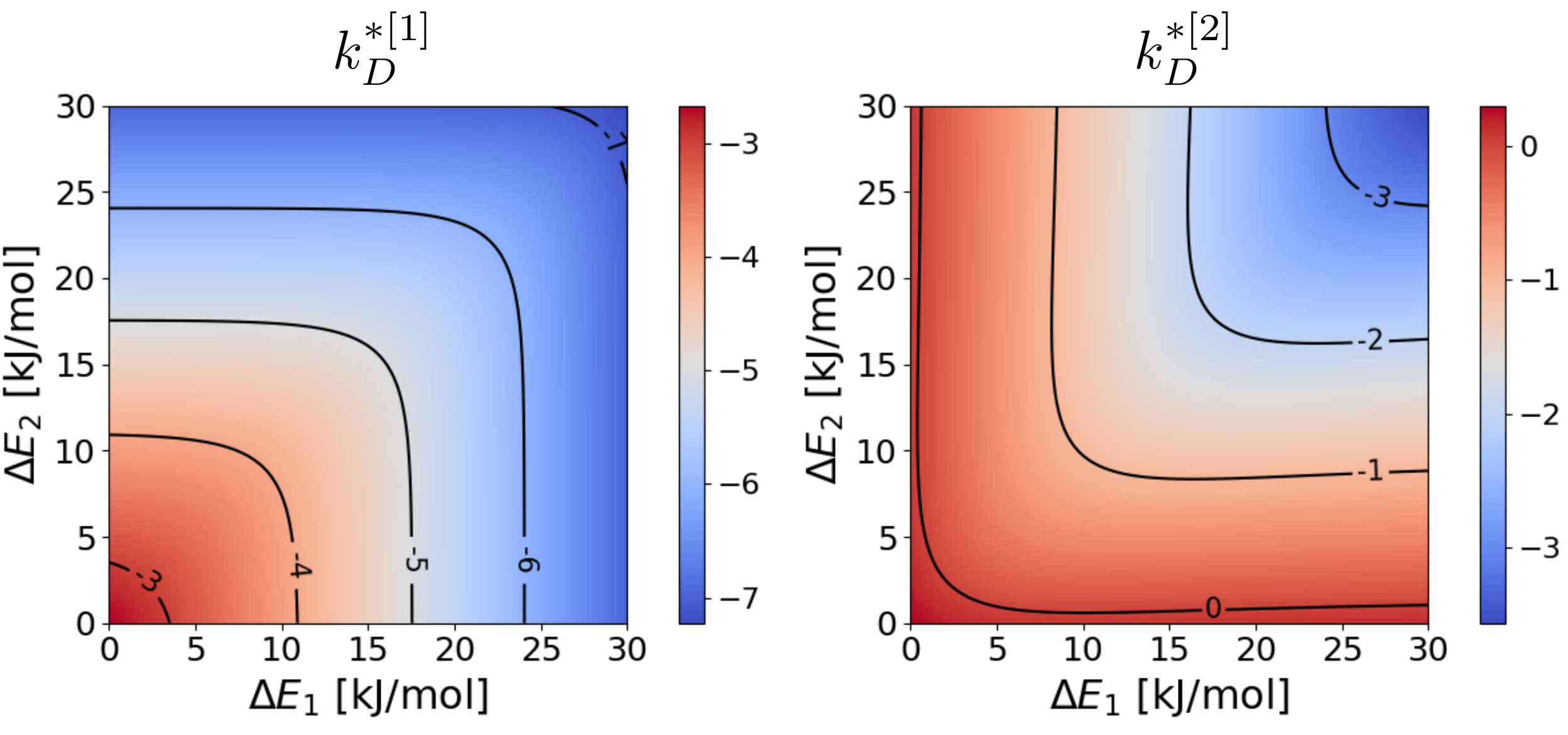}
\caption{{\bf The mobility rates that mark stabilisation crossovers of furanose are well separated 
and depend markedly on the energy barriers}.
Density plot of $k_D^{\ast[2]}$ and $k_D^{\ast[1]}$ (in units of $\mu$) 
in the plane $\Delta E_1,\Delta E_2$ in logarithmic (base 10) scale.
Parameters as in Fig.~\ref{f:ultrastab} (see Table~I in the main text)}
\label{f:k12starmaps} 
\end{figure}

\noindent As transport of reactants  across the temperature gradient is put back into the picture,
non-equilibrium effects start shifting the steady state away from thermal equilibrium, 
as sustained currents set in, coupling mass transport to chemical transformations. 
Fig.~\ref{f:ultrastab} shows clearly that, when production of furanose proceeds faster 
than pyranose (i.e. $\Delta E_2 > \Delta E_1$), this causes the population of 
furanose to grow beyond its highest equilibrium value at values of Da of order one. 
We refer to this non-equilibrium effect as {\em ultra-stabilisation} of furanose. \\
\indent More precisely, with the choice of parameters derived from our NMR experiments, 
the steady-state solutions~\eqref{e:pssgen}  display three regimes as a function of the 
transport rate $k_D$, identified by two distinct crossovers occurring 
at characteristic rates $k_D^{\ast[1]}$ and $k_D^{\ast 2} \geq k_D^{\ast[1]}$,
as illustrated in Fig.~\ref{f:ultrastab}. These can be computed straightforwardly from 
Eq.~\eqref{e:pssgen} and show a marked dependence on the 
choice of the energy barriers $\Delta E_1$ and $\Delta E_2$, as illustrated by Fig.~\ref{f:k12starmaps}.
However, with the choice of parameters extracted from 
our equilibrium experiments, it turns out that the two crossovers are always distinct,
$k_D^{\ast[2]}$ being at least three orders of magnitude larger than $k_D^{\ast[1]}$ irrespective 
of the magnitude of the energy barriers. \\
%
%
%
\indent Increasing $k_D$ past $k_D^{\ast[1]}$,
the steady-state populations of same-species states 
become independent of the temperature gradient.
In this regime, the molar fractions at the cold side become identical to  the values at the hot side
and the two boxes can be regarded as a well-stirred open chemical reactor.
This is due to a small, but non-zero steady mass current that  
circulates in the direction $0\to1\to2\to3\to4\to5\to0$ 
(see Fig.~\ref{f:3statesmodel}) and hence only 
involves transport of closed-ring species 
across the temperature gradient in the hot-to-cold direction.
Fig.~\ref{f:k12starmaps} (left) shows that this first non-equilibrium effect occurs
in a regime where  mass transport  is still much slower than chemical reactions, 
i.e. for values of Da $ = \mu/k_D > 10^3$. Our calculations show that 
sustained currents that stabilise the high-energy 
species kick in at exponentially slower transport rates as the energy barriers increase.  
More precisely, it can be shown that (see the following)
\begin{equation}
\label{e:k1sth}
k_D^{\ast[1]} \simeq \frac{\ds \mu}{\ds P^{\rm eq}_{E_1}(T_1) 
                                \left[
                                   f_1 \left( 
                                         e^{\Delta E_1/k_BT_1} + e^{\Delta E_2/k_BT_1}
                                       \right) + 
                                   f_2 \left( 
                                         e^{\Delta E_1/k_BT_2} + e^{\Delta E_2/k_BT_2}
                                       \right)  
                                \right]}
\end{equation}
where $f_i = P^{\rm eq}_{E_2}(T_i)/P^{\rm eq}_{E_0}(T_i)$.
In other words, when the energy barriers are small, hence chemical steps faster, 
molecular species need to diffuse or be advected faster to settle into a different 
steady state than at thermal equilibrium. \\
%
%
\indent When the mobility rate increases past $k_D^{\ast[1]}$, 
a second crossover to a different steady state is seen to occur at $k_D \simeq  k_D^{\ast[2]} $.
We find that, when the two energy barriers are well separated, i.e.
$\Delta E_1 \ll \Delta E_2$ or $\Delta E_2 \ll \Delta E_1$, 
this second crossover corresponds to a Damk\"{o}hler number Da$^\ast = \mu/k_D^{\ast[2]} \simeq \mathcal{O}(1)$,
provided the smaller barrier does not exceed the average thermal energy provided at the hot end, $k_BT_2$. 
This is the fast-transport regime, where transport and chemical reaction time scales match 
(see Fig.~\ref{f:k12starmaps}, right). The relation between the critical value of the Damk\"{o}hler number 
and the main determinants of relaxation kinetics, i.e. the energy barriers, 
can be encapsulated in a remarkably transparent formula, notably
\begin{equation}
\label{e:k2sTHE}
\text{Da}^\ast = \frac{1}{\ds e^{-\Delta E_1/k_BT_M} + e^{-\Delta E_2/k_BT_M} }                
\end{equation}
where we have indicated with $T_M = (T_1 + T_2)/2$ the average temperature  of the system.
From Eq.~\eqref{e:k2sTHE} it can be readily seen that the requirement for kinetic selection, i.e. 
scenarios where one of the barriers is much higher than the average temperature
and the other much lower, indeed correspond to the timescale-matching condition Da$^\ast = \mathcal{O}(1)$.\\
\indent The second crossover is seen to occur at the same value of Da 
irrespective of whether production of furanoses is faster than of pyranoses or viceversa. 
However,  the physical consequences of this second transition turn out to be strikingly divergent,
depending on which of the two barriers is the largest and on their relative magnitude. 
If furanose production is the fastest relaxation channel, 
i.e. $\Delta E_1 \ll \Delta E_2$, the second crossover leads to ultra-stabilisation of the 
most unstable closed form of the sugar~\footnote{Conversely, if $\Delta E_1 \gg \Delta E_2$,
pyranose production production proceeds much faster, and it is that species which 
is ultra-stablised.}, as sustained currents keep the 
population of furanose  beyond the highest value accessible at equilibrium,  $P^{\rm eq}_{E_1}(T_2)$.
It can be shown that the relative furanose-to-pyranose population 
takes a remarkably simple expression in the ultra-stabilisation regime
\begin{equation}
\label{e:Rexpl}
\frac{P_{E_1}^\infty}{P_{E_2}^\infty} \simeq
\frac{P_{E_1}^{\rm eq}(T_2)}{P_{E_2}^{\rm eq}(T_2)}  e^{(E_0-E_1)\Delta T/k_BT_1T_2} 
\end{equation}
%
%
where $\Delta T = T_2-T_1$ is the temperature gradient. 
Eq.~\eqref{e:Rexpl} conveys an important piece of physical information. 
The relative dissipation-sustained enrichment of the high-energy furanose species, $E_1$, 
is magnified (exponentially) the larger its energy 
separation from the high-energy  intermediate (linear) state, $E_0$.
Of course, the larger the temperature gradient, the greater this effect. \\
%
%
%
%
%
\noindent In general, it not difficult to check that the condition for 
the  two-crossover trend observed in Fig.~\ref{f:ultrastab} to exist is
\begin{equation}
\label{e:k12codn}
A_1 \gg \sqrt{A_0A_2}
\end{equation}
In this case, the functions $x_i$ will display a three-plateau trend 
as follows
\begin{equation}
\label{e:x13regimes}
x_i(k_D) \simeq
\begin{cases}
\frac{\ds\alpha_{0i}}{\ds A_0} & \text{for} \quad k_D \ll k_D^{\ast[1]} \\
\frac{\ds\alpha_{1i}}{\ds A_1} \stackrel{\rm def}{=} x^{\rm int}_i & \text{for} 
\quad k_D^{\ast[1]} \ll k_D \ll k_D^{\ast[2]}\\
\frac{\ds\alpha_{2i}}{\ds A_2} & \text{for} \quad k_D \gg k_D^{\ast[2]} 
\end{cases}
\end{equation}
where the two crossover mobility rates are given by  
\begin{equation}
\label{e:kstar12}
k_D^{\ast[1]} = \frac{A_0}{A_1} \qquad 
k_D^{\ast[2]} = \frac{A_1}{A_2} 
\end{equation}
With the choice of parameters 
extracted from our NMR experiments, the second crossover corresponds to a transport 
rate that is at least 3 orders of magnitude larger than that marking the first crossover
(see Fig.~\ref{f:k12starmapswTHE}).\\
\indent Furthermore, with the following definitions, which mark the 
relevant populations that characterise the three different regimes ({\rm int} for intermediate regime),  
\begin{eqnarray}
&&x^{\rm eq}_i  \stackrel{\rm def}{=} \lim_{k_D\to\,0}  x_i(k_D) = \frac{\ds\alpha_{0i}}{\ds A_0} \label{e:xeqdef}\\
&&x^{\rm int}_i \stackrel{\rm def}{=} \frac{\ds\alpha_{1i}}{\ds A_1} \label{e:xintqdef} \\
&&x^{\infty}_i \stackrel{\rm def}{=} \lim_{k_D\to\,\infty}  x_i(k_D) = \frac{\ds\alpha_{2i}}{\ds A_2} \label{e:xinfdef}
\end{eqnarray}
the steady-state populations, Eq.~\eqref{e:pssgen}, can be cast in the more transparent form
(see again Eq.~\eqref{e:kstar12})
\begin{equation}
\label{e:xismart}
x_i(k_D) = \frac{\ds x_i^\infty k_D^2 + x^{\rm int}_i k_D^{\ast[2]} 
                 k_D + x^{\rm eq}_i  k_D^{\ast[1]}k_D^{\ast[2]}}
                {k_D^2 + k_D^{\ast[2]} k_D +  k_D^{\ast[1]}k_D^{\ast[2]}}
\end{equation}
%

\section{Analytical expressions for the two crossover rates}

An analytical approximation of the crossover rates $k_D^{\ast[1]}$ and $k_D^{\ast[2]}$ 
can be obtained by noticing that the same crossovers are observed 
for the molar fractions of the same species in either box as functions of 
the transport rate $k_D$. Hence, the abscissae 
of the (logarithmic) inflection points can be estimated by considering ratios 
between same-species molar fractions. 

\subsection{The first crossover}

In order to compute $k_D^{\ast[1]}$, it is expedient to 
focus on the high-energy species (nodes 1 and 2 in the graph, corresponding 
to $T_2$ and $T_1$, respectively). We have
\begin{equation}
\label{e:x1ovx2}
\frac{x_1}{x_2} = \frac{\alpha_{21} k_D^2 + \alpha_{11} k_D + \alpha_{01}}
                       {\alpha_{22} k_D^2 + \alpha_{12} k_D + \alpha_{02}}
\end{equation}                       
In the region of the first crossover we can neglect the terms of order $k_D^2$ 
in Eq.~\eqref{e:x1ovx2}, so that 
\begin{equation}
\label{e:x1ovx2ap1}
\frac{x_1}{x_2} \approx \frac{\alpha_{11} k_D + \alpha_{01}}
                             {\alpha_{12} k_D + \alpha_{02}}
\end{equation}                       
from which we can estimate $k_D^{\ast[1]}$ as
\begin{equation}
\label{e:k1sdef}
k_D^{\ast[1]} = \frac{\alpha_{02}}{\alpha_{12}}
\end{equation}                       
After some algebraic manipulation, we have
\begin{equation}
\label{e:k1sTHE}
k_D^{\ast[1]} = \frac{\mu}{P_{E_1}^{\rm eq}(T_1)[Q(T_1) + Q(T_2)]}
\end{equation}
where 
\begin{equation}
\label{e:Q}
Q(T) = \frac{P_{E_2}^{\rm eq}(T)}{P_{E_0}^{\rm eq}(T)} 
\left( e^{\Delta E_1/k_BT} + e^{\Delta E_2/k_BT}\right)
\end{equation}
%
%
%
%

%
\begin{figure}[t!]
\centering
\includegraphics [width=12truecm]{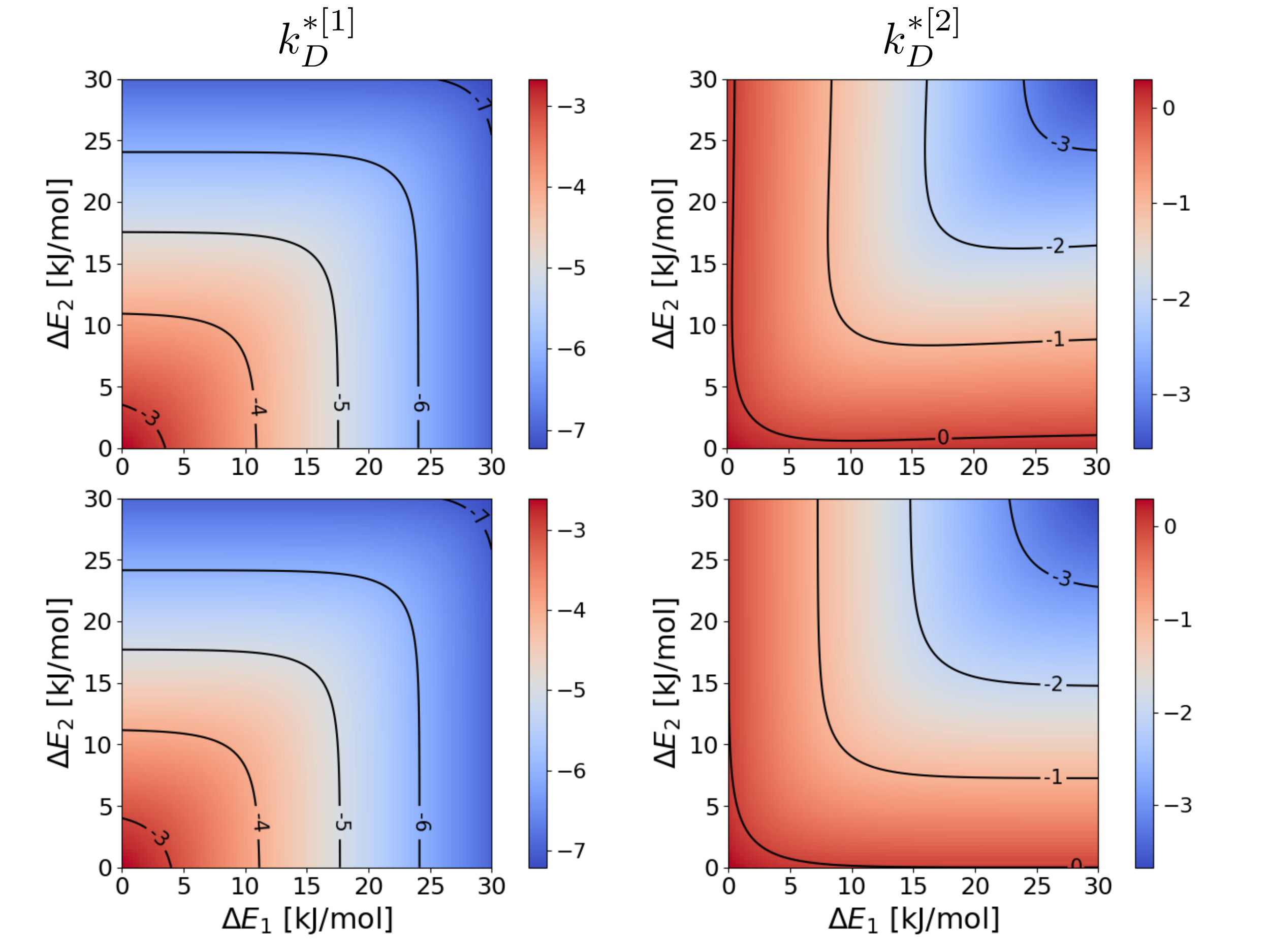}
\caption{
Density plot of $k_D^{\ast[2]}$ and $k_D^{\ast[1]}$ in the plane $\Delta E_1,\Delta E_2$
in logarithmic (base 10) scale. The top panels correspond to the exact expressions given 
by~ Eq.\eqref{e:kstar12}. The bottom panel are density plots of the explicit 
expressions, Eq.~\eqref{e:k1sTHE} ($k_D^{\ast[1]}$) and Eq.~\eqref{e:k2sTHE} ($k_D^{\ast[2]}$).
Parameters are $\eta_1 = 29.34$, $\eta_2 = 1.945$, $E_0=19$ kJ/mol, $E_1=13.6$ kJ/mol, 
$E_2=3.1$ kJ/mol, corresponding to the average values for the two furanose and pyranose
enantiomers measured from equilibrium NMR experiments (see main body).}
\label{f:k12starmapswTHE} 
\end{figure}
%

\subsection{The second crossover}

In this case, it is expedient to focus on the low-energy species instead 
(nodes 5 and 4 in the graph, corresponding  to $T_2$ and $T_1$, respectively).
Thus, we have
\begin{equation}
\label{e:x5ovx4}
\frac{x_5}{x_4} = \frac{\alpha_{25} k_D^2 + \alpha_{15} k_D + \alpha_{05}}
                       {\alpha_{24} k_D^2 + \alpha_{14} k_D + \alpha_{04}}
\end{equation}                       
In the region of the second crossover we can neglect the zero-order terms 
in Eq.~\eqref{e:x5ovx4}, so that 
\begin{equation}
\label{e:x1ovx2ap2}
\frac{x_1}{x_2} \approx \frac{\alpha_{25} k_D + \alpha_{15}}
                             {\alpha_{24} k_D + \alpha_{14}}
\end{equation}                       
from which we can estimate $k_D^{\ast[2]}$ as
\begin{equation}
\label{e:k2sdef}
k_D^{\ast[2]} = \frac{\alpha_{14}}{\alpha_{24}}
\end{equation}
The full expression in terms of the transition rates reads (see again the partial 
diagrams in Fig.~\ref{f:3statesmodel})
\begin{equation}
\label{e:kdsfull}
k_D^{\ast[2]} = \frac{k_{23}k_{34}(k_{01}+k_{05}) + 
                    k_{10}k_{05}(k_{34}+k_{32}) + 
                    k_{10}k_{23}(k_{05}+k_{34}) + 
                    k_{50}k_{34}(k_{10}+k_{23})}{k_{23}k_{05} + k_{10}k_{34}}
\end{equation}
A useful analytical expression can be derived by considering the limit $\eta_1 \to \infty$, 
which is seen to be a good approximation already for $\eta_1 > 0.1$.  
Recalling expressions~\eqref{e:rates}, after some algebra, we get
\begin{equation}
\label{e:kdslimeta1}
k_D^{\ast[2]} = \mu\frac{G(T_1,T_2)F(T_1,T_2) + G(T_2,T_1)F(T_2,T_1)}{G(T_1,T_2) + G(T_2,T_1)}
\end{equation}
where 
\begin{eqnarray}
\label{e:GF}
G(T_1,T_2) &=& \frac{P_{E_0}^{\rm eq}(T_1)}{P_{E_1}^{\rm eq}(T_1)}e^{-\Delta E_1/T_1 -\Delta E_2/T_2} \\
F(T_1,T_2) &=& e^{-\Delta E_2/T_1} 
               \left( 
               1 + e^{(\Delta E_2 - \Delta E_1)/T_2} 
               \right) 
\end{eqnarray}
It turns out that formula~\eqref{e:kdslimeta1} has an extremely simple and transparent 
approximation, namely
\begin{eqnarray}
\label{e:k2sTHE}
k_D^{\ast[2]} &\simeq& \frac{\mu}{2} \left(e^{-\Delta E_1/k_BT_1} + e^{-\Delta E_2/k_BT_2} +
                                         e^{-\Delta E_2/k_BT_1} + e^{-\Delta E_1/k_BT_2}\right) \nonumber \\
              &=&  \mu \left(
                              e^{-\Delta E_1/k_BT_M} + e^{-\Delta E_2/k_BT_M}
                        \right) + \mathcal{O}(\Delta T^2)                    
\end{eqnarray}
where we have indicated with $T_M= (T_1+T_2)/2$ the average temperature  of the system and with
$\Delta T = T_2-T_1$ the temperature gradient.    
In figure~\ref{f:k12starmapswTHE} we compare the explicit analytical 
expressions, Eq.~\eqref{e:k1sTHE} and Eq.~\eqref{e:k2sTHE}, with the 
exact values of the crossover rates, computed through Eqs.~\eqref{e:kstar12}.
It is seen that the two analytical estimates are in excellent agreement with 
the results of the full calculations, including the extremely simple 
formula~\eqref{e:k2sTHE}.

\begin{figure}[ht!]
\centering
\includegraphics[width=12truecm]{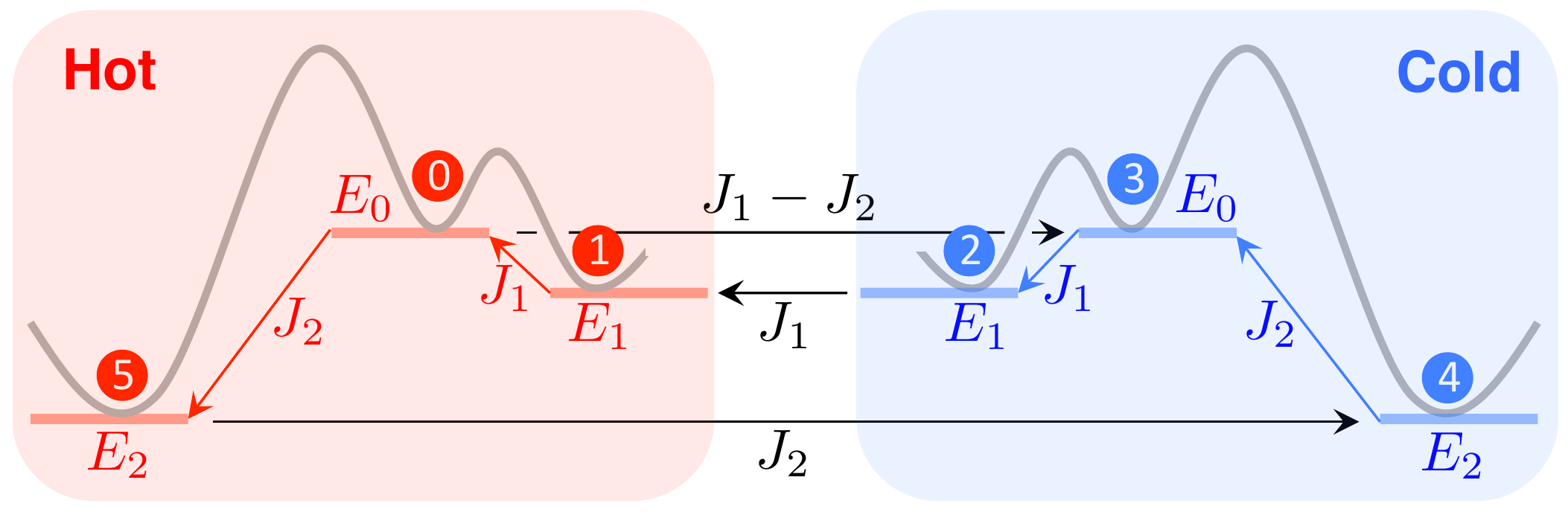}
\includegraphics [width=7truecm]{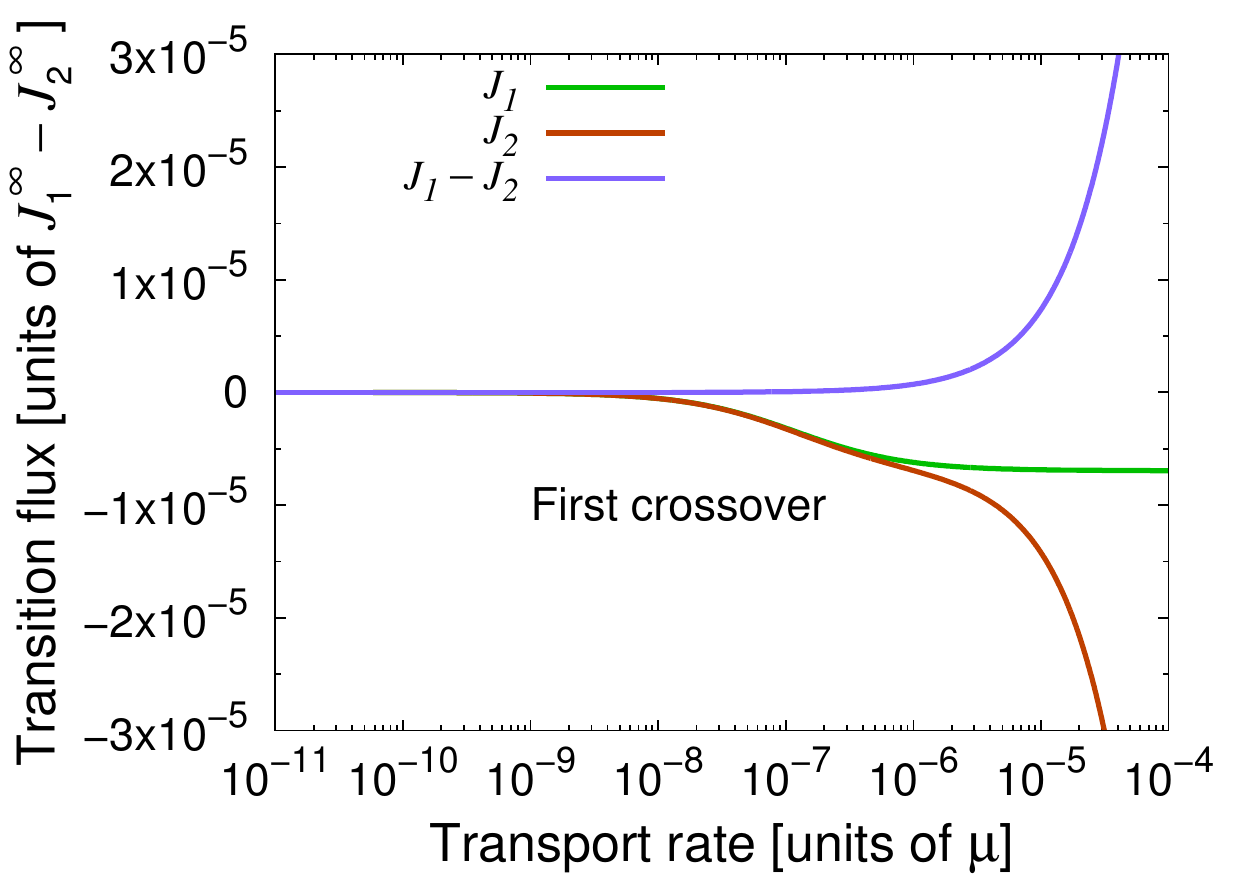}
\includegraphics [width=7truecm]{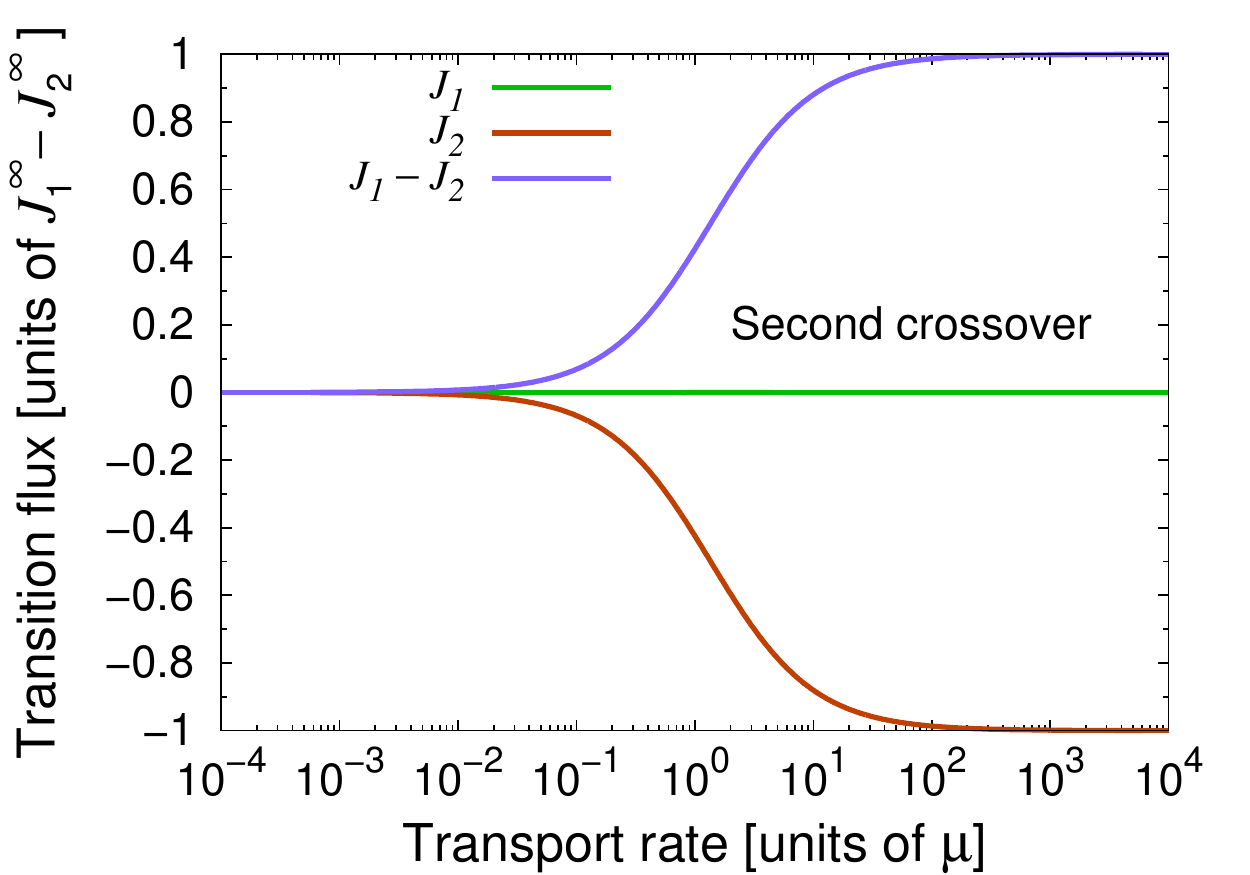}
\includegraphics [width=7truecm]{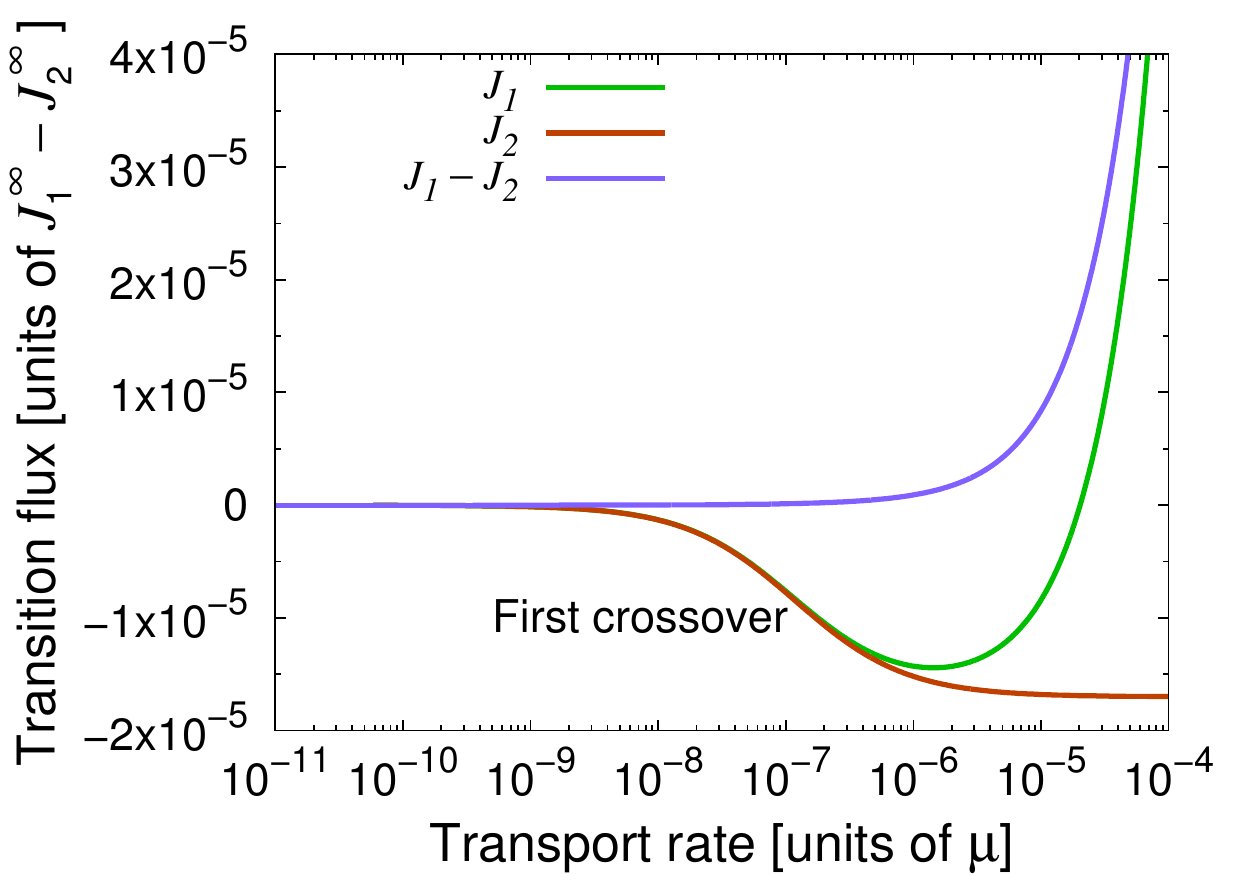}
\includegraphics [width=7truecm]{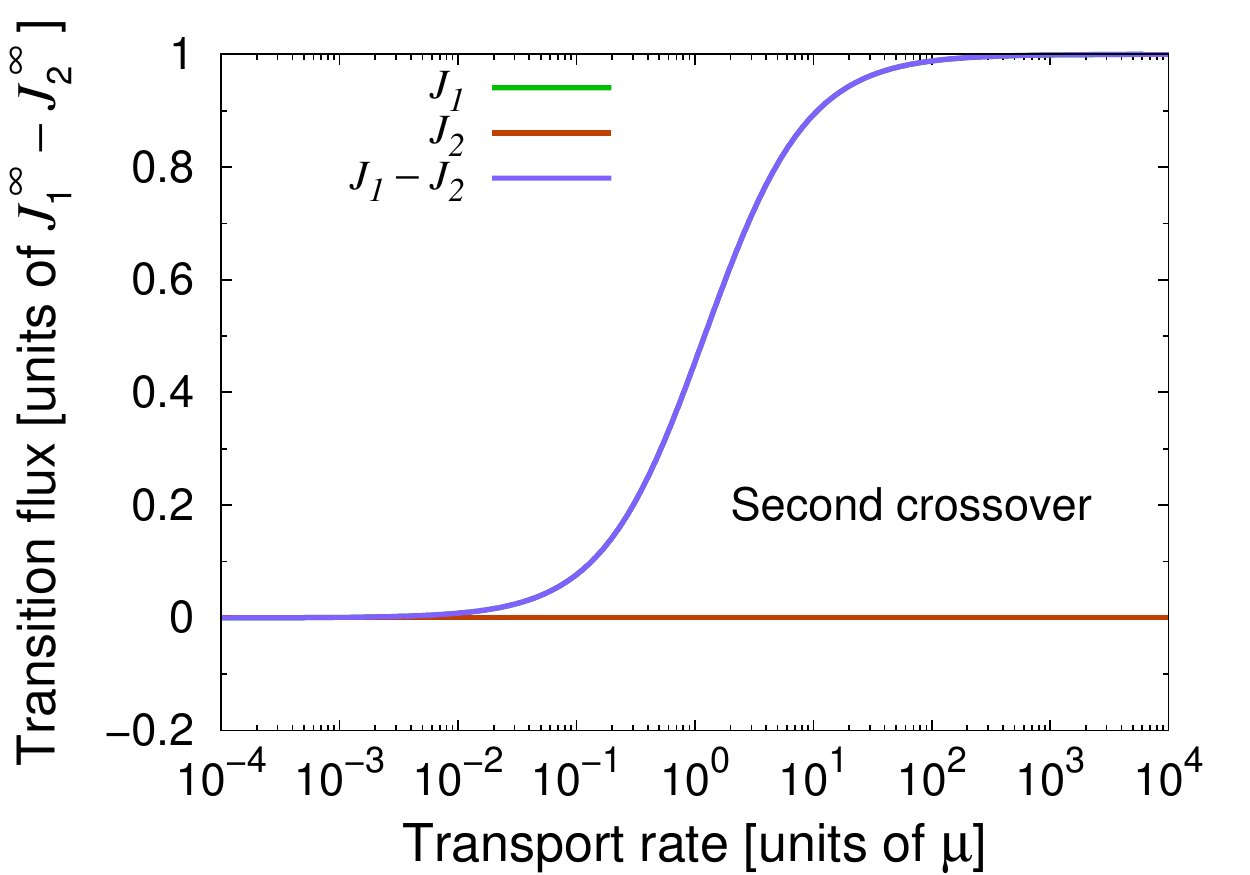}
\caption{\textbf{The steady-state populations are sustained by steady currents, which 
circulate in different subgraphs in the first and second crossover regimes.} 
Fluxes as a function of the mobility rate.
Top graphs: ultra-destabilisation scenario, $\Delta E_1 = 30$ kJ/mol, $\Delta E_2 = 0$. 
A steady clockwise flux in the $b$ cycle ($J_1-J_2 =  -J_2>0$) sustains as $k_D\to\infty$ 
a ultra-high population of the low-energy state $E_2$ and an ultra-low population of the 
high-energy state $E_1$.
Bottom graphs, ultra-stabilisation scenario, $\Delta E_1 = 0$, $\Delta E_2 = 30$ kJ/mol. 
In this case, steady counterclockwise flux in the $a$ cycle ($J_1-J_2 = J_2>0$) sustains as $k_D\to\infty$ 
an ultra-low population of the low-energy state $E_2$ and a ultra-high population of the 
high-energy state $E_1$.
Clockwise from top to bottom: 
$\Delta E_1 = 0$ and varying $\Delta E_2 \in [0,30]$ kJ/mol.
$\Delta E_1 = 10$ kJ/mol and varying $\Delta E_2 \in [0,30]$ kJ/mol
(the blue dashed line corresponds to $\Delta E_2= \Delta E_1 = 10$ kJ/mol).
$\Delta E_2 = 0$ kJ/mol and varying $\Delta E_1 \in [0,30]$ kJ/mol.
Varying $\Delta E_1 = \Delta E_2 \in [0,30]$ kJ/mol.
Parameters are $\eta_1 = 29.34$, $\eta_2 = 1.945$, $E_0=19$ kJ/mol, $E_1=13.6$ kJ/mol, 
$E_2=3.1$ kJ/mol, corresponding to the average values for the two furanose and pyranose
enantiomers measured from equilibrium NMR experiments.}
\label{f:fluxes} 
\end{figure}
%

\section{Cycle fluxes}

\begin{figure}[t!]
\centering
\includegraphics [width=10.5truecm]{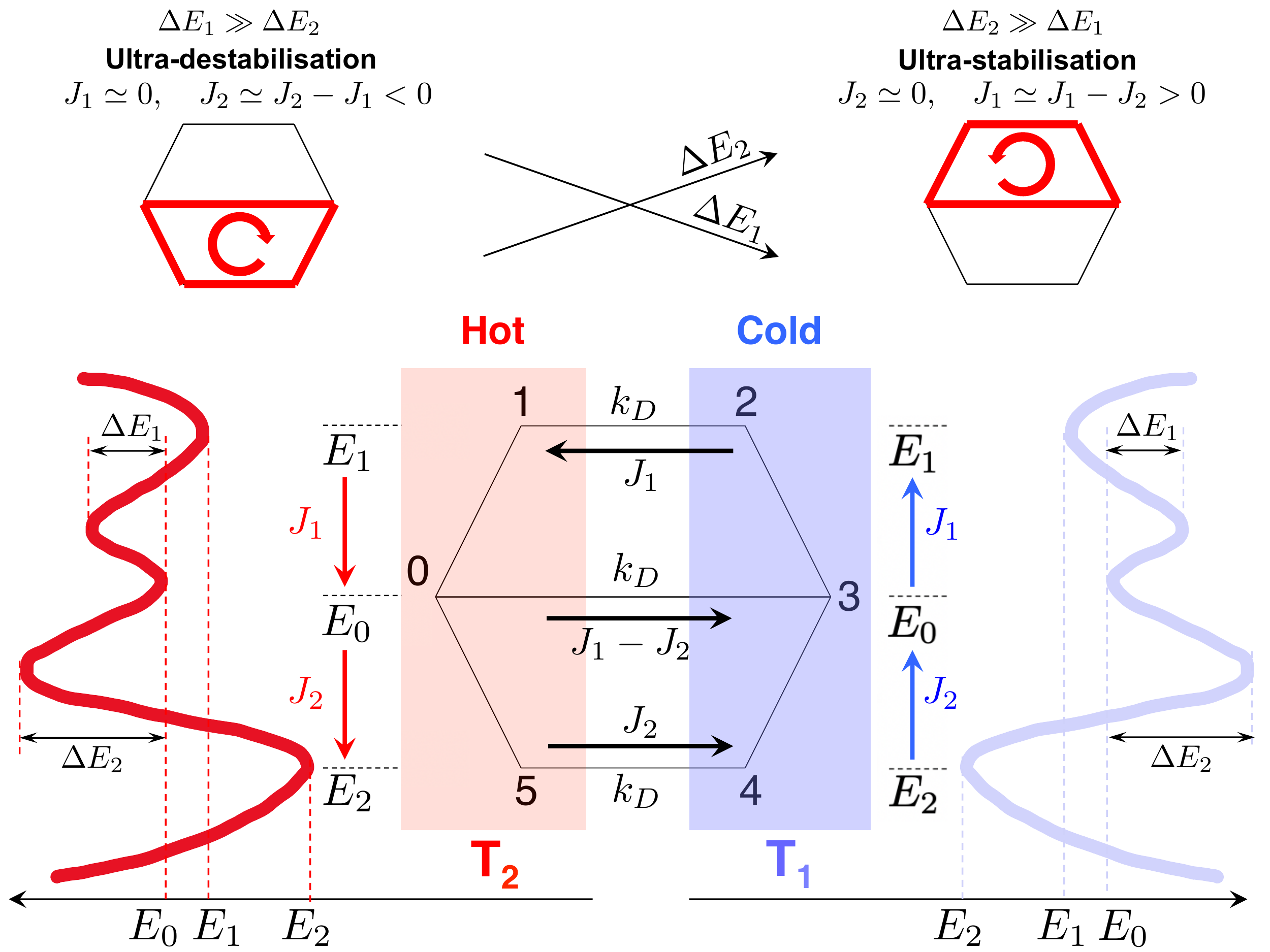}
\includegraphics [width=3.5truecm]{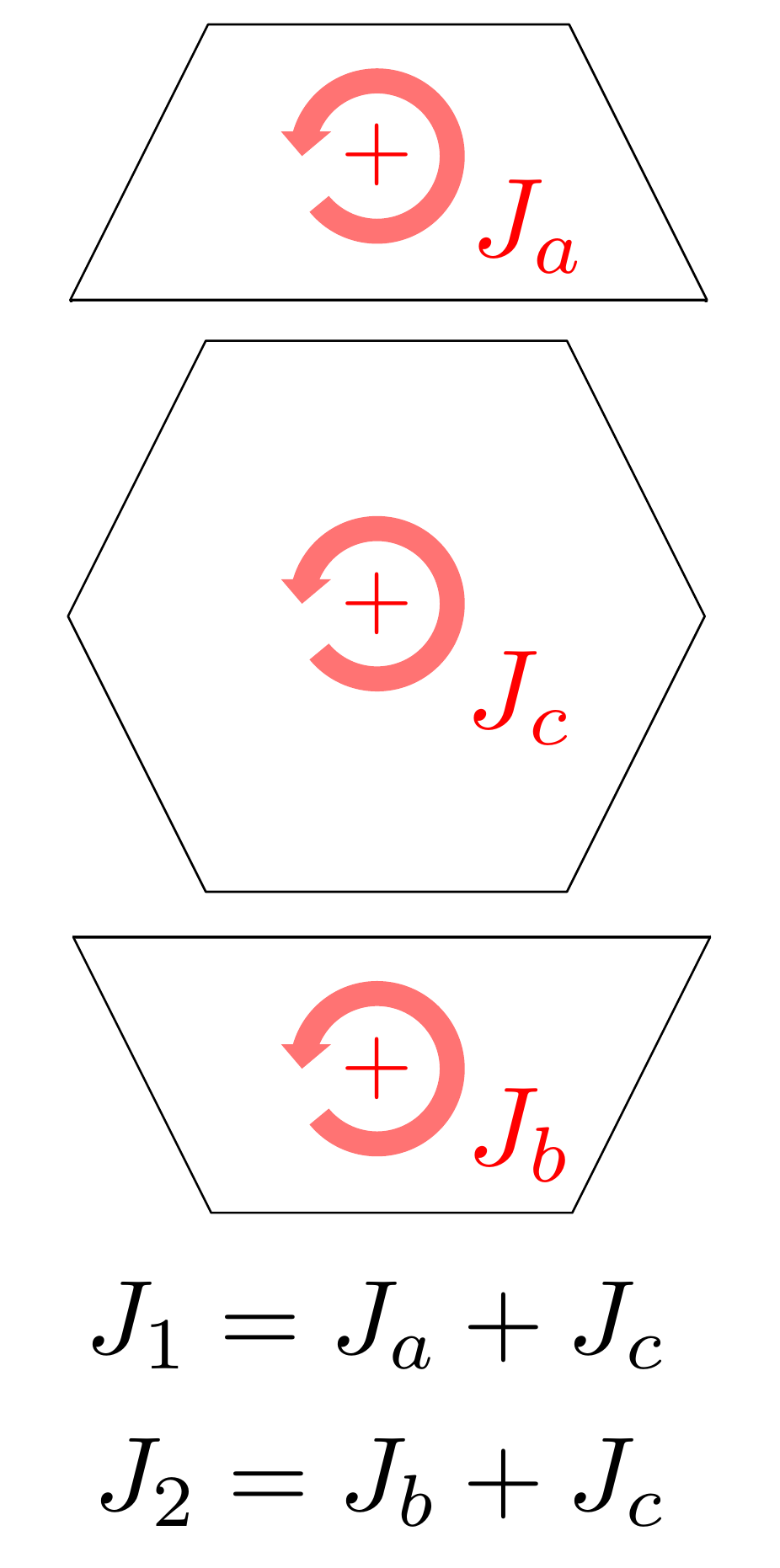}
\includegraphics [width=10.5truecm]{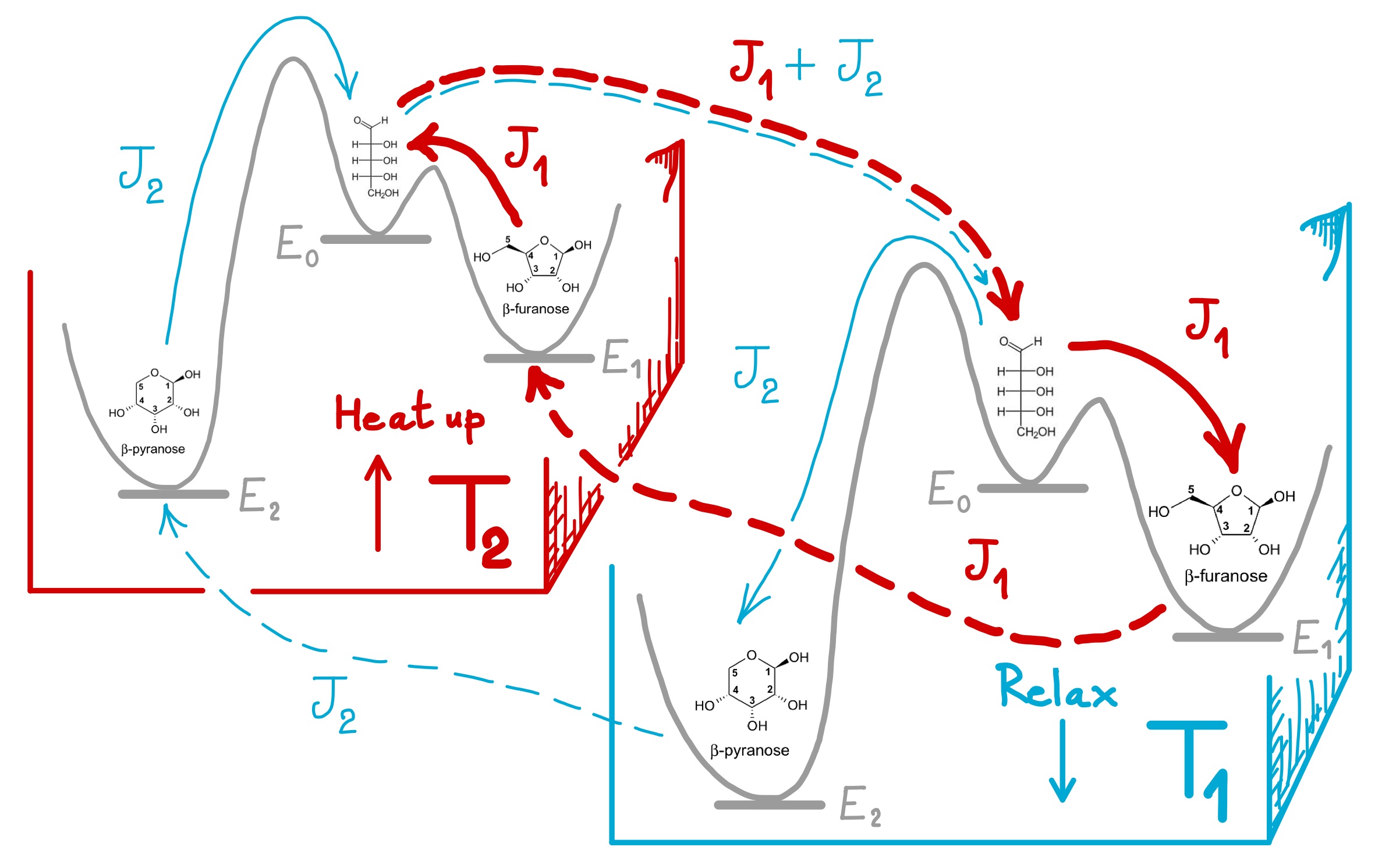}
\caption{{\bf Coarse-grained ribose isomerisation 
network in a steady temperature gradient}. Top: as one of the linear-to-cycle transition 
becomes faster than the other, the corresponding barrier becoming smaller, 
its population is boosted as a result of sustained non-equilibrium currents, as shown 
within the corresponding sub-graphs.
Bottom left: scheme of the network with the corresponding energy landscapes and  
transition fluxes. The latter can be conveniently expressed in terms of two independent 
linear combinations of the three cycle fluxes $J_a$, $J_b$ and $J_c$
(right).
The direction of the transition fluxes shown reflects the choice made for their 
definition, that is, an arrow from $i$ to $j$ stands for the $i\to j$  flux
(which can be either positive or negative). Bottom. 
Illustration of the steady system of currents that circulate in the system 
in the case of ultra-stabilisation, $\Delta E_2 \gg \Delta E_1$. Dashed arrows
denote transport, solid lines stand for chemical transformations.
The current $J_2$ is much smaller than $J_1$ in the fast transport limit, 
$k_D \gg k_D^{\ast[2]}$. For example, for $\Delta E_2 =$ 30 kJ/mol and $\Delta E_1 = 0$,
$J_2/J_1 \simeq 6.5\times 10^{-5}$, while for  
$\Delta E_2 =$ 20 kJ/mol and $\Delta E_1 = 5$ kJ/mol,
$J_2/J_1 \simeq 1.2\times 10^{-2}$}
\label{f:graphfluxes} 
\end{figure}

With reference to Fig.~\ref{f:3statesmodel}, there are three cycle fluxes in our system. According to the
general diagrammatic prescription described in Ref.~\cite{Hill:1989aa}, the ratio between the one-way 
cycle fluxes (according to the chosen convention for positive fluxes) can be computed rather 
straightforwardly. More precisely, one has
\begin{eqnarray}
\frac{J_{a+}}{J_{a-}} &=& \frac{k_{32}k_{10}}{k_{23}k_{01}} = e^{(E_0-E_1)/k_BT_m}\label{e:Jafrac}\\
\frac{J_{b+}}{J_{b-}} &=& \frac{k_{43}k_{05}}{k_{34}k_{50}} = e^{-(E_0-E_2)/k_BT_m}\label{e:Jbfrac}\\
\frac{J_{c+}}{J_{c-}} &=& \frac{k_{43}k_{32}k_{10}k_{05}}{k_{34}k_{23}k_{01}k_{50}} = 
                          e^{-(E_1-E_2)/k_BT_m}\label{e:Jcfrac}
\end{eqnarray}
where 
\begin{equation}
\label{e:Tm}
\frac{1}{T_m} = \frac{1}{T_1} - \frac{1}{T_2} 
\end{equation}
\begin{eqnarray}
&&J_a = J_{a+} - J_{a-} =  J_{a-} \left[
                                       e^{(E_0-E_1)/k_BT_m} - 1 
                                \right] \geq 0       \label{e:Ja}\\
&&J_b = J_{b+} - J_{b-} = J_{b-} \left[ 
                                       e^{-(E_0-E_2)/k_BT_m} - 1 
                                \right] \leq 0       \label{e:Jb}\\
&&J_c = J_{c+} - J_{c-} = J_{c-} \left[  
                                       e^{-(E_1-E_2)/k_BT_m} - 1
                                 \right] \leq 0      \label{e:Jc}
\end{eqnarray}
The above inequalities follow directly from the strict positivity of one-way cycle fluxes, 
which are rational functions of products of rates~\cite{Hill:1989aa}.\\
\indent Much insight can be gained by inspecting the {\em transition} fluxes, $J_{ij} = -J_{ji}$, between 
any two neighboring states in the diagram. These are simply given by the sum (including the appropriate 
sign) of cycle fluxes for those cycles that comprise the link $ij$ (from $i$ to $j$). 
For example, the flux between the nodes representing the linear species from hot to cold is 
$J_{03} = J_a - J_b$. It should be noted that, at least in principle, transition fluxes are 
observable, while cycle fluxes are not.\\
\indent In general, if a network comprises $N$ nodes and $M$ links, there are 
$N_J = M - N + 1$ independent transition fluxes. In our case, $M = 7$, $N = 6$, hence
$N_J = 2$.  These can be conveniently identified as 
\begin{eqnarray}
&&J_1 = J_a + J_c \label{e:J1}\\
&&J_2 = J_b + J_c \label{e:J2}
\end{eqnarray}
so that all transition fluxes are determined as $J_{32} = J_{21} = J_{10} = J_1$, 
$J_{05} = J_{54} = J_{43} = J_2$, $J_{03} = J_1 - J_2$. Note that the second law of 
thermodynamics is obviously not violated in the non-equilibrium steady state, 
as the net energy flux still proceeds from the hot to the cold box. 
Recalling the definition of transition fluxes and Eqs.~\eqref{e:J1},~\eqref{e:J2}, 
the total heat flux from the hot source to the cold one reads
\begin{eqnarray}
\label{e:2ndprinciple}
\dot{Q} &=&  E_0 J_{03} + E_1 J_{12} + E_2 J_{54} \nonumber\\
                 &=&  (E_0-E_1)J_1 - (E_0-E_2)J_2 \nonumber\\
                 &=&  (E_2-E_1)J_2 + (E_0-E_1)(J_1-J_2) \geq 0
\end{eqnarray}
where the last passage follows from the definitions of cycle fluxes~\eqref{e:Ja},~\eqref{e:Jb} 
and~\eqref{e:Jc}, which imply $J_2 \leq 0$, $J_1-J_2\geq 0$.
The currents circulating in the system in the limit $k_D\to\infty$ are illustrated  in 
Fig.~\ref{f:graphfluxes}. \\

%
%

\section{Rate of entropy production and energy dissipation}

It is interesting to compute the rate of entropy production and to relate it to the 
non-equilibrium stabilisation sustained by steady currents in the system, as illustrated 
in the main text. The rate of entropy production can be computed as~\cite{Schnakenberg:1976aa}
\begin{equation}
\label{e:Sdot}
\dot{S} = \frac{1}{2} \sum_{i,j=0}^5 J_{ji} 
          \log 
                \left( 
                     \frac{k_{ji} x_j}{k_{ij} x_i}
                \right)
\end{equation}
where $J_{ji} = k_{ji} x_j - k_{ij} x_i$ is the transition flux from node $j$ to node $i$.
Recalling Eqs.~\eqref{e:Jafrac},~\eqref{e:Jbfrac} and~\eqref{e:Jcfrac} and taking into account 
the definitions of the operational fluxes $J_1$ and $J_2$, Eqs.~\eqref{e:J1} and~\eqref{e:J2}, 
it is not difficult to see that Eq.~\eqref{e:Sdot} simplifies to
\begin{equation}
\label{e:Sdotsimple}
\dot{S} = \frac{1}{k_B T_m} 
                        \left[ J_1 (E_0-E_1) - J_2(E_0-E_2) \right] 
\end{equation}
where $T_m$ is the reduced temperature defined by Eq.~\eqref{e:Tm}. Direct inspection of 
Eq.~\eqref{e:2ndprinciple}, shows that the rate of entropy production is proportional, 
as it should, to the net heat flux flowing from the hot reservoir to the cold one, namely 
\begin{equation}
\label{e:2ndprincipleSdot}
k_B\dot{S} = \left( 
                  \frac{1}{T_1} - \frac{1}{T_2}
          \right) \dot{Q}
\end{equation}
From the analysis of fluxes reported in the previous section, it is not difficult to
realise that (see again Fig.~\ref{f:3statesmodel}) and Fig.~5 in the main text)
in general the following equalities hold (at any time)
\begin{eqnarray}
\label{e:J1J2expl}
J_1 &=& k_D(x_2 - x_1) \nonumber \\
J_2 &=& k_D(x_5 - x_4) 
\end{eqnarray}
Using Eqs.~\eqref{e:xismart} and observing that $\alpha_{22}=\alpha_{21}$ and
$\alpha_{25}=\alpha_{24}$ (see again the topology of the spanning trees shown 
in Fig.~\ref{f:3statesmodel}), we get
\begin{eqnarray}
\label{e:J1J2expl_rates}
J_1 &=& k_D \frac{\ds (x^{\rm int}_2 - x^{\rm int}_1) k_D^{\ast[2]} k_D + (x^{\rm eq}_2-x^{\rm eq}_1)  k_D^{\ast[1]}k_D^{\ast[2]}}
                {k_D^2 + k_D^{\ast[2]} k_D +  k_D^{\ast[1]}k_D^{\ast[2]}}  \nonumber \\
J_2 &=& k_D \frac{\ds (x^{\rm int}_5 - x^{\rm int}_4) k_D^{\ast[2]} k_D + (x^{\rm eq}_5-x^{\rm eq}_4)  k_D^{\ast[1]}k_D^{\ast[2]}}
                {k_D^2 + k_D^{\ast[2]} k_D +  k_D^{\ast[1]}k_D^{\ast[2]}}
\end{eqnarray}
Plugging Eqs.~\eqref{e:J1J2expl_rates} in the definition of the rate of 
entropy production, Eq.~\eqref{e:Sdotsimple}, the latter can be cast in the particularly 
transparent form
\begin{equation}
\label{e:Sdot_transp}
\dot{S} = \frac{\ds \dot{S}_\infty k_D^2 + \dot{S}_{\rm int}k_Dk_D^{\ast[2]}}
               {\ds k_D^2 + k_D^{\ast[2]} k_D +  k_D^{\ast[1]}k_D^{\ast[2]}}
               \simeq
               \begin{cases}
               \dot{S}_{\rm int} \frac{\ds k_D}{\ds k_D + k_D^{\ast[1]}} & \text{for} \quad k_D \ll k_D^{\ast[2]} \\
               \frac{\ds \dot{S}_\infty k_D + \dot{S}_{\rm int}k_D^{\ast[2]}}{\ds k_D + k_D^{\ast[2]}}  &
               \text{for} \quad k_D \gg k_D^{\ast[1]}
               \end{cases}
\end{equation}
In particular, we get the following simple asymptotic forms,
\begin{equation}
\label{e:Sdot_limits}
\dot{S} \simeq 
\begin{cases}
               \dot{S}_{\rm int} \left( 
                                    \frac{\ds k_D}{\ds  k_D^{\ast[1]}}
                                 \right)  & \text{for} \quad k_D \to \, 0 \\
               \dot{S}_\infty \left[ 1 
                              - \frac{\ds k_D^{\ast[2]}}{\ds k_D}
                              \left( 
                                     1 -
                                     \frac{\ds  \dot{S}_{\rm int}}{\ds  \dot{S}_\infty }
                              \right)
                               \right] &
               \text{for} \quad k_D \to \, \infty
               \end{cases}
\end{equation}
%

\subsection{The limit of vanishing transport rate}

\begin{figure}[t!]
\centering
\includegraphics[width=12truecm]{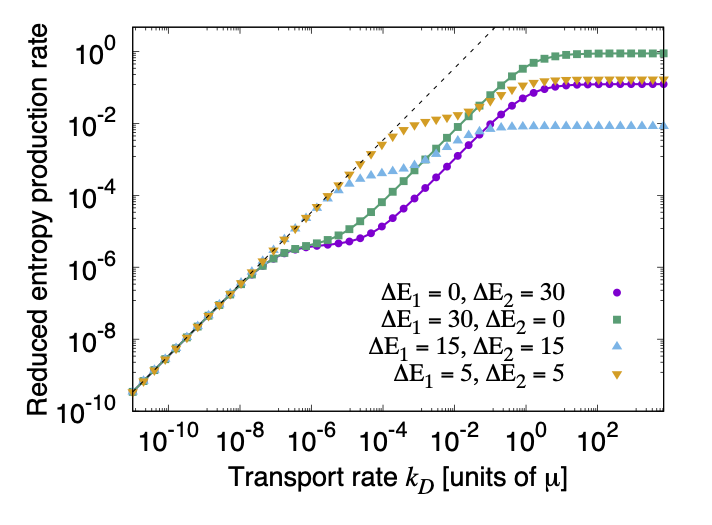}
\caption{\label{f:Sdotfit} Plot of the reduced entropy production rate for different choices 
of the two barriers $\Delta E_1$ and $\Delta E_2$. The solid lines are fits with the
expression~\eqref{e:Sdot_transp}. The vanishing transport prediction~\eqref{e:Sdotkd0}
is shown explicitly as a dashed line. All other parameters are fixed as described in 
the main text at the values determined experimentally for the ribose isomerisation 
network.}
\end{figure}

\noindent In the limit $k_D\to0$, we see from Eq.~\eqref{e:J1J2expl} that the two operational fluxes 
read 
\begin{eqnarray}
\label{e:J1J2explkd0}
J_1 &\simeq& \frac{1}{2} k_D \left( 
                              P_{E_1}^{\rm eq}(T_1) - P_{E_1}^{\rm eq}(T_2)
                             \right)  \nonumber \\
J_2 &\simeq& \frac{1}{2} k_D \left( 
                              P_{E_2}^{\rm eq}(T_2) - P_{E_2}^{\rm eq}(T_1)
                             \right) 
\end{eqnarray}
where we have used Eqs.~\eqref{e:xieqexpl}. Hence, we have from expression~\eqref{e:Sdotsimple}
\begin{equation}
\label{e:Sdotkd0}
\dot{S} = \frac{1}{k_B T_m} 
                        \left[ 
                             (E_0-E_1) 
                             \left( 
                                P_{E_1}^{\rm eq}(T_1) - P_{E_1}^{\rm eq}(T_2)
                             \right) - 
                             (E_0-E_2)
                             \left( 
                                P_{E_2}^{\rm eq}(T_2) - P_{E_2}^{\rm eq}(T_1)
                             \right)
                        \right ] k_D
\end{equation}                        
We see that the entropy production rate increases linearly with the transport rate $k_D$
at small values of the latter. The expression~\eqref{e:Sdotkd0} can be simplified 
further and made more transparent by expanding it in powers of $\Delta T/T_M$, 
where $T_M = (T_1+T_2)/2$ ($\approx 0.85$ with the choice of temperatures considered in this 
work, and recalling that $P_{E_0}^{\rm eq}(T_i) \ll P_{E_1}^{\rm eq}(T_i),P_{E_2}^{\rm eq}(T_i)$, 
$i=1,2$. After a straightforward calculation, we get
\begin{equation}
\label{e:Sdotkd0DT0}
\dot{S} = \frac{1}{2} P_{E_1}^{\rm eq}(T_M) P_{E_2}^{\rm eq}(T_M) 
         \left( 
            \frac{E_2 - E_1}{k_BT_M}
         \right)^2
         \left(
           \frac{\Delta T}{T_M}
         \right)^2 k_D + \mathcal{O}(k_D^2,\Delta T^3)
\end{equation}
Fig.~\ref{f:Sdotfit} shows the trend of the entropy production rate 
$\dot{S}$ as a function of the transport rate $k_D$ 
for different choices of the barriers $\Delta E_1$ and $\Delta E_2$.
It can be appreciated that the system dissipates 
more the faster the transport, until a maximum dissipation rate is reached 
that depends only on the choice of the barriers and the imposed temperature gradient.

%
%
%
%

%
%

\providecommand{\latin}[1]{#1}
\makeatletter
\providecommand{\doi}
  {\begingroup\let\do\@makeother\dospecials
  \catcode`\{=1 \catcode`\}=2 \doi@aux}
\providecommand{\doi@aux}[1]{\endgroup\texttt{#1}}
\makeatother
\providecommand*\mcitethebibliography{\thebibliography}
\csname @ifundefined\endcsname{endmcitethebibliography}
  {\let\endmcitethebibliography\endthebibliography}{}

%
\end{document}